\documentclass[preprint]{jpsj2} 
\title{Ground State Properties and Optical Conductivity of the Transition Metal Oxide ${\rm Sr_{2}VO_{4}}$}

\author{Yoshiki \textsc{Imai}$^{1,2,3}$
\thanks{e-mail: imai@phy.saitama-u.ac.jp}
 and Masatoshi \textsc{Imada}$^{1,2,4}$}

\inst{$^{1}$Institute for Solid State Physics, University of Tokyo, Kashiwanoha, Kashiwa, Chiba, 277-8581\\
$^{2}$PRESTO, Japan Science and Technology Agency \\
$^{3}$Department of Physics, Saitama University, Saitama, 338-8570 \\
$^{4}$Department of Applied Physics, University of Tokyo, Hongo, Bunkyo-ku, Tokyo, 113-8656.}
\abst{
Combining first-principles calculations with a technique for many-body problems, we investigate properties of the transition metal oxide ${\rm Sr_{2}VO_{4}}$ from the microscopic point of view.
By using the local density approximation (LDA), the high-energy band structure is obtained, while screened Coulomb interactions are derived from the constrained LDA and the GW method. The renormalization of the kinetic energy is determined from the GW method. By these downfolding procedures, an effective Hamiltonian at low energies is derived.
Applying the path integral renormalization group method to this Hamiltonian, we obtain ground state properties such as the magnetic and orbital orders. Obtained results are consistent with experiments within available data.
We find that ${\rm Sr_{2}VO_{4}}$ is close to the metal-insulator transition.
Furthermore, because of the coexistence and competition of ferromagnetic and antiferromgnetic exchange interactions in this system, an antiferromagnetic and orbital-ordered state with a nontrivial and large unit cell structure is predicted in the ground state.
The calculated optical conductivity shows characteristic shoulder structure in agreement with the experimental results. This suggests an orbital selective reduction of the Mott gap.
}

\kword{${\rm Sr_{2}VO_{4}}$, local density approximation, first-principles calculation, path-integral renormalization group, spin-orbital coupled order, orbital degeneracy}

\begin{document}
\maketitle

\section{Introduction}

The discovery of high temperature superconductivity in cuprates has stimulated interest to strongly correlated electron systems,
such as transition metal oxides and heavy fermion compounds. These materials exhibit a variety of magnetic, dielectric, and
transport properties, which are caused by the strong Coulomb interactions between conduction electrons\cite{imad98}.

\begin{figure}[b]
\begin{center}
\includegraphics[width=8cm]{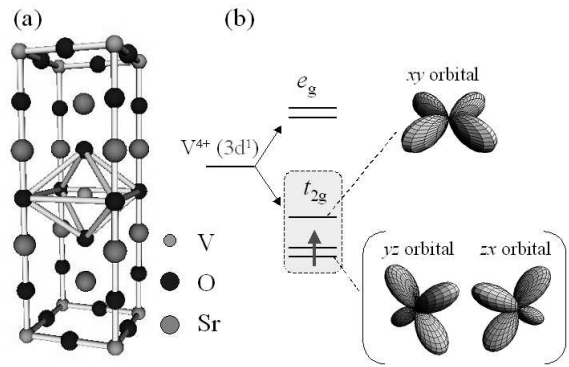}
\end{center}
\caption{(a) Crystal structure of ${\rm Sr_{2}VO_{4}}$ and (b) $3d$ levels. }
\label{lattice}
\end{figure}

${\rm Sr_{2}VO_{4}}$ is one of the typical transition metal oxides isostructural to the parent compound of the high-$T_{\rm c}$
superconducting material, ${\rm La_{2}CuO_{4}}$, which has ${\rm K_{2}NiF_{4}}$-type structure (shown in Fig.\ref{lattice}(a)).
Namely, this compound has a layered perovskite structure with a strong two-dimensional anisotropy. The similarity (or duality)
of the electronic configuration between ${\rm V^{4+}}$ (one $3d$ electron) in ${\rm Sr_{2}VO_{4}}$ and ${\rm Cu^{2+}}$ (one $3d$ hole)
in ${\rm La_{2}CuO_{4}}$ should be noted. However, this duality is not complete. The crystal-field effect splits $3d$ orbitals of
${\rm V^{4+}}$ into $e_{g}$ and $t_{2g}$ orbitals (see Fig.\ref{lattice}(b)). Furthermore the strong two-dimensionality splits $t_{2g}$ orbitals into the $xy$
orbital and degenerate $yz$ and $zx$ orbitals where the orbital degeneracy of $d^{1}$ electron remains between $yz$ and $zx$ orbitals,
in contrast to the orbitally non-degenerate structure of the cuprates, where $d_{x^2-y^2}$ orbital is isolated near the Fermi level.
In the cuprates, in addition, the oxygen $p$ orbital is strongly hybridized while this hybridization is small in ${\rm Sr_{2}VO_{4}}$.
The crystal field splitting between $xy$ orbital and others ($yz$ and $zx$) are also rather small ($\sim 0.08$eV in the LDA calculation) in
${\rm Sr_{2}VO_{4}}$ and the $xy$ orbital is also expected to play some role in the ground state.
%
%
This comparison suggests that studies on ${\rm Sr_{2}VO_{4}}$ may contribute in revealing the role of orbital degeneracy in comparison to
the cuprates.  It may also provide us with insights on the physics and mechanism of the high-$T_{\rm c}$ superconductors.

Motivated by the analogy with ${\rm La_{2}CuO_{4}}$, there exist several experimental studies on
${\rm Sr_{2}VO_{4}}$\cite{cyro90,rey90,mats90,noza91,gian95,mats05}. Recently thin film on ${\rm LaSrAlO_{4}}$
became available and investigated\cite{mats03,mats05}. The electronic resistivity shows semiconducting behavior,
whereas the resistivity does not diverge at low temperature \cite{noza91,gian95}. The activation energy observed
in the temperature dependence of the resistivity lies in the range of 50-90 meV \cite{noza91} and 70 meV \cite{mats03},
which indicates that the ${\rm Sr_{2}VO_{4}}$ is a Mott insulator with a very small Mott gap.  In fact metallic phase easily
appears by La doping\cite{mats03,mats05}. The magnetic susceptibility follows the Curie-Weiss law at temperature above 100K.
With decreasing temperature, cusps appear at 45K\cite{noza91} and at 10K\cite{cyro90}. Although the value of N{\'e}el
temperature strongly depends on samples, the ground state appears to possess an antiferromagnetic (AF) order.
Because these experimental results were obtained mostly from powder samples except for thin film results~\cite{mats03,mats05},
it is hard to draw conclusions on the intrinsic properties.  As far as we know, studies on orbital orders are not available
in the literature. The optical conductivity measured for thin films show a characteristic shoulder structure~\cite{mats05}, which requires a careful consideration because it is not easily explained from the Hartree-Fock picture. 

To get insight into physics of Sr$_2$VO$_4$ and predict intrinsic properties of this compound in an accurate computational framework,
we apply a newly developed computational method proposed as a hybrid first-principles approach combined with the low-energy solver for
the low-energy effective model~\cite{imai05}.
For the low-energy solver, we employ the path-integral renormalization group (PIRG) method~\cite{imad00,kash01,mori02,wata03,mizu04}.
As we will remark later, more conventional theoretical approaches predict rather controversial results: The local density approximation (LDA) concludes paramagnetic metal while the Hartree-Fock approximation (HFA) suggests the ferromagnetic insulator.
In the present paper, we show that these two results are not accurate and the careful considerations on the strong electron correlation
effect and quantum fluctuations are crucially important in analyzing the properties of this compound. The present first-principles PIRG
calculation predicts that this compound is very close to the Mott transition in the insulating side with a tiny Mott gap.  In addition,
it predicts that the spin and orbital are strongly coupled with frustration effects, leading to a nontrivial and long-period ordering
structure of spins and orbitals.  We also show that the optical conductivity is consistent with the experimental results and is interpreted by the orbital selective reduction of the Mott gap.  A part of the present results has already been reported in a letter~\cite{imai05}.

This paper is organized as follows. In the next section, we derive the Hubbard type effective Hamiltonian from the results of the first-principles calculation. In \S3 and \S4, we refer to the low-energy effective Hamiltonian and the conventional approaches. In \S5, the formalism of PIRG method and its application are briefly described. We show results by HFA and PIRG and discuss physical properties of ${\rm Sr_{2}VO_{4}}$ in \S6. A brief summary and discussions are given in \S7.

\section{High-Energy Electronic Structure and Downfolding}
\label{High-Energy}
Since electrons far from the Fermi level are not usually affected by strong fluctuations arising from correlation effects, the LDA and GW approximation offers reasonable and
suitable methods in this regime. We perform elimination of the high-energy degrees of freedom by a downfolding procedure by the density
functional approach.  We then derive the effective low energy Hamiltonian~\cite{solo04,solo05a,solo05b}, which can be treated by the model
Hamiltonian approach. We employ a downfolding procedure, in which the LDA based on the LMTO basis functions is combined with the GW approach
and the constrained LDA method to derive the screened Coulomb interaction and the renormalized band dispersion.

Since global band structure obtained by LDA calculations shows that $3d$ $t_{2g}$ orbitals are close to
the Fermi level and are rather isolated from other orbitals (Fig.~\ref{ldados}), it allows to construct the effective low-energy Hamiltonian by extracting
only the degrees of freedom for the V $3d$ $t_{2g}$ in ${\rm Sr_{2}VO_{4}}$. Since the crystal field splitting between $xy$ orbital
and other orbitals is quite small ($\sim$ 0.08eV in LDA calculation), $xy$ orbital should not be neglected. We formulate the downfolding procedure in detail in this section.

\subsection{Kinetic-Energy Part}
\label{Kinetic}

  The kinetic-energy part of the model Hamiltonian has been derived
using the downfolding method,\cite{solo04,solo05b}
starting from the electronic structure
in LDA.

  For practical purposes we use the linear-muffin-tin-orbital (LMTO)
method supplemented with the atomic spheres approximation (ASA).\cite{LMTO}
The LMTO basis functions, $\{ | \chi \rangle \}$ (the
so-called muffin-tin orbitals -- MTOs) have many similarities with
the orthogonalized atomic orbitals. They are constructed from
solutions of Kohn-Sham (KS) equations inside atomic spheres, $\phi$, calculated
at certain energies $E_\nu$ (typically, the center of
gravity of occupied band or of the entire band), and their
energy derivatives $\dot{\phi}$.
At the atomic spheres boundaries,
$\phi$ and $\dot{\phi}$ match continuously
and differentiably onto certain envelope function (typically,
irregular solutions of Laplace equation, which rapidly decay in the
real space). Since $\langle \phi |
\dot{\phi} \rangle$$=$$0$, the basis is orthogonal
and $\{ | \chi \rangle \}$
can be regarded as the Wannier functions for the full KS Hamiltonian $H_{\rm KS}$.

%
\begin{figure}[tb]
\begin{center}
\includegraphics[width=8cm]{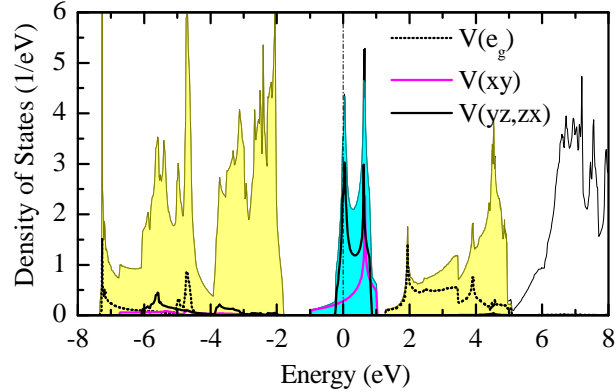}
\end{center}
\caption{(color online) LDA density of states. The energy is measured form the Fermi level (0 ev).  }
\label{ldados}
\end{figure}
%
\begin{figure}[b]
\begin{center}
\includegraphics[width=8cm]{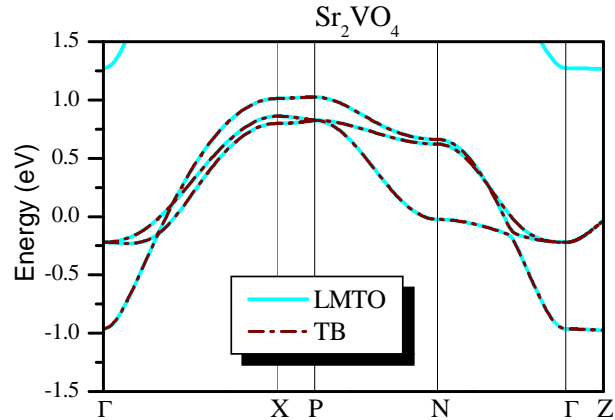}
\end{center}
\caption{(color online) The band structure of V 3$d$ $t_{2g}$ orbitals for ${\rm Sr_{2}VO_{4}}$. Solid lines represent the result of LMTO,  while dashed-dotted lines stand for our tight binding approximation.}
\label{band}
\end{figure}
%

  In order to describe properly the electronic structure of Sr$_2$VO$_4$
in the valent part of the spectrum, we
used 67 MTOs (including those associated with the
empty spheres, which have been added in order to improve the atomic
spheres approximation for the loosely packed
layered perovskite
structure -- see Table~\ref{tab.LMTOparameters}). The
KS Hamiltonian in the basis of these MTOs will be denoted by
$\hat{H}$$\equiv$$\langle \chi | H_{\rm KS} | \chi \rangle$.
\begin{table}[h!]
\caption{Atomic positions (units of $a$$=$$3.837$\AA and
$c$$=$$12.576$\AA), atomic radii (in \AA) and basis functions
included in LMTO calculations of Sr$_2$VO$_4$. The symbol `Em'
stands for the empty spheres. The crystal structure is shown in
Fig.~\protect\ref{lattice}.} \label{tab.LMTOparameters}
\begin{tabular}{ccccc}  \hline
type of atom  & position & atomic radius & LMTO basis & number of atoms \\  \hline
Sr     &  $(0,0,0.354)$          &  $1.924$  & $5s5p4d4f$  &  $2$   \\
V      &  $(0,0,0)$              &  $1.383$  & $4s4p3d$    &  $1$   \\
O$_1$  &  $(0.5,0,0)$            &  $0.988$  & $2s2p$      &  $2$   \\
O$_2$  &  $(0,0,0.158)$          &  $1.012$  & $2s2p$      &  $2$   \\
Em$_1$ &  $(0.5,0,$-$0.25)$      &  $0.648$  & $1s$        &  $2$   \\
Em$_2$ &  $(0.313,0,$-$0.208)$   &  $0.437$  & $1s$        &  $8$   \\ \hline
\end{tabular}
\end{table}
Corresponding densities of states and the
energy dispersion of the $t_{2g}$ bands
are shown
in Figs.~\ref{ldados} and \ref{band}, respectively.
We note an excellent agreement with calculations employing
more accurate full-potential linearized-augmented-plane-wave (FLAPW)
method.\cite{sing91}

  What we want to do next is to describe
the behavior of three $t_{2g}$ bands located near the Fermi level
by certain tight-binding (TB) Hamiltonian $\hat{h}$, which, contrary
to $\hat{H}$, is formulated in basis of
only three (unknown yet) Wannier orbitals centered at each V site.

  By denoting the subspace spanned by three MTOs of $t_{2g}$ type as
$\{ | d \rangle \}$ and the rest of MTOs as $\{ | r \rangle \}$,
such that $\{ | \chi \rangle \}$$=$$\{ | d \rangle
\}$$\oplus$$\{ | r \rangle \}$, the equations for eigenvalues
and eigenfunctions of the KS Hamiltonian $\hat{H}$
can be rearranged identically as
\begin{eqnarray}
( \hat{H}^{dd}-\omega ) | d \rangle  +  \hat{H}^{dr} | r \rangle & = & 0, \label{eqn:seceq1}\\
\hat{H}^{rd} | d \rangle  +  ( \hat{H}^{rr}-\omega ) | r \rangle & =
& 0. \label{eqn:seceq2}
\end{eqnarray}
By eliminating $| r \rangle$ from eq.~(\ref{eqn:seceq2})
and substituting it in eq.~(\ref{eqn:seceq1}) one obtains
the effective $\omega$-dependent Hamiltonian in the basis of
$d$-states
$$
\hat{H}^{dd}_{\rm eff}(\omega) = \hat{H}^{dd} - \hat{H}^{dr}
(\hat{H}^{rr} - \omega)^{-1}\hat{H}^{rd}
$$
and the ''overlap'' matrix
$$
\hat{S}(\omega)=1+\hat{H}^{dr}
(\hat{H}^{rr}-\omega)^{-2}\hat{H}^{rd},
$$
such that $\langle d | \hat{S} | d \rangle$$=$$1$. Then, the
required TB Hamiltonian, $\hat{h}$ is obtained after the
orthonormalization of the vectors $| d \rangle$$\rightarrow$$|
\tilde{d} \rangle$$=$$\widehat{S}^{1/2}| d \rangle$ and fixing the
energy $\omega$ in the center of gravity of $t_{2g}$ band ($\omega_0$):
\begin{equation}
\hat{h} = \hat{S}^{-1/2}(\omega_0) \hat{H}^{dd}_{\rm
eff}(\omega_0) \hat{S}^{-1/2}(\omega_0). \label{eqn:TB}
\end{equation}

  The downfolding method is nearly perfect and well reproduces the behavior
of three $t_{2g}$ bands in the reciprocal (${\bf k}$) space (Fig.~\ref{band}).
The obtained Hamiltonian is then Fourier transformed into
the real space. The site-diagonal elements of $\hat{h}$$=$$\| h_{ij}^{mm'} \|$,
obtained after such a transformation,
describe the crystal-field (CF) splitting,
whereas the off-diagonal elements stand for
transfer integrals.
Here $i$ and $j$ stand for atomic sites while $m,m'$ represent the band indices.

\subsection{Wannier Functions}

  The Wannier
functions for the isolated $t_{2g}$ band can be reconstructed from
the parameters of the kinetic energy
obtained in the downfolding method.\cite{solo05b}
In order to do so, we solve an inverse problem and search for the functions
$\{ W_i^m \}$, which after applying to the KS Hamiltonian in the real space,
generate the matrix elements
$h_{ij}^{mm'}$$=$$\langle W_i^m | H_{\rm KS} | W_j^{m'} \rangle$.
The method has been described in details in the previous publication.\cite{solo05b}
The basic idea is to search $W_i^m$ in the form of MTO:\cite{LMTO}
$$
| W_i^m \rangle \propto |\phi_i^m \rangle + \sum_{j,m'}
|\phi_j^{m'} \rangle (h_{ji}^{m'm} - E_\nu \delta_{ij} ),
$$
with subsequent orthonormalization and orthogonalization to other bands.

\begin{figure}[tb]
\begin{center}
\includegraphics[width=6cm]{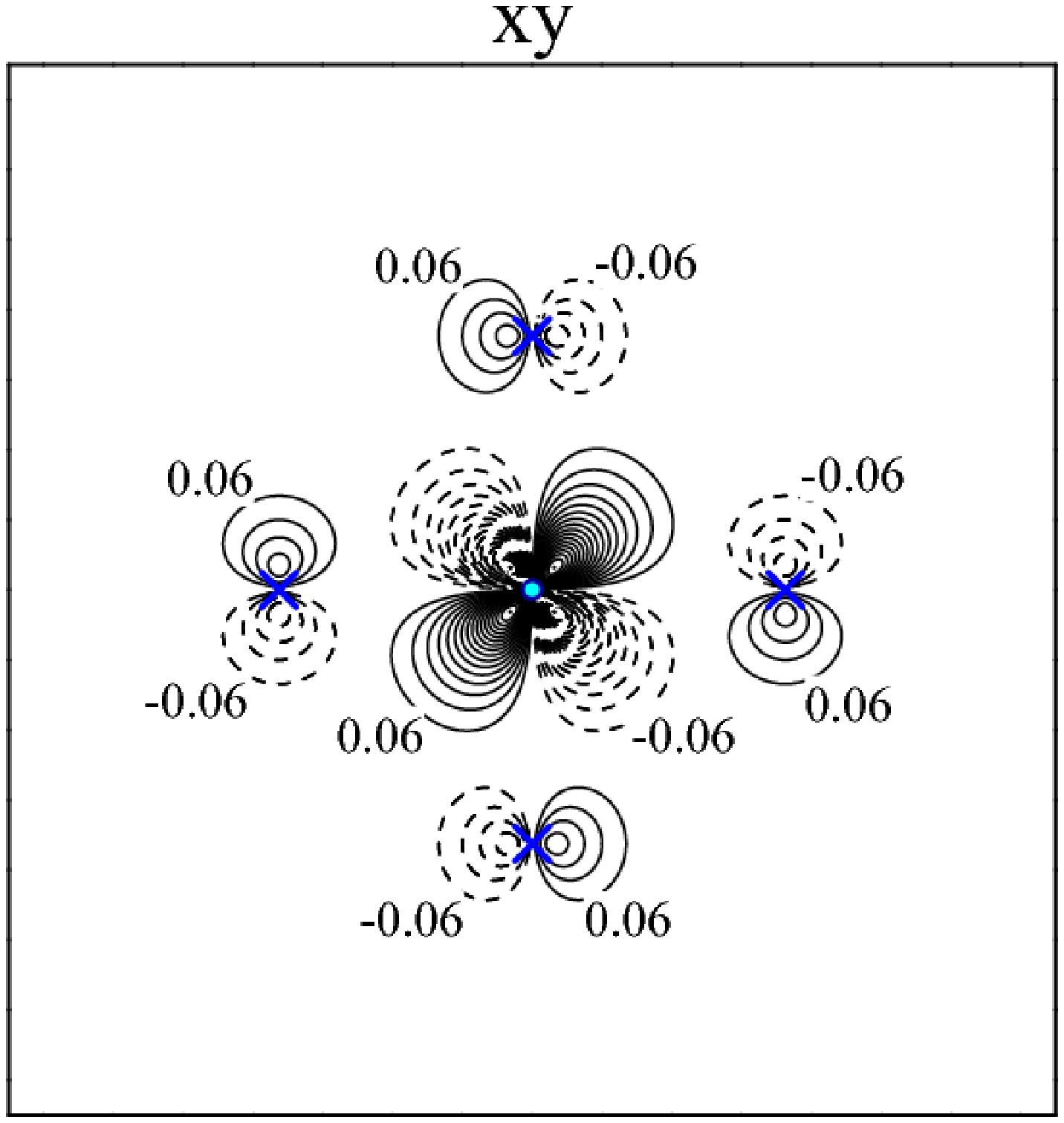} \includegraphics[width=6cm]{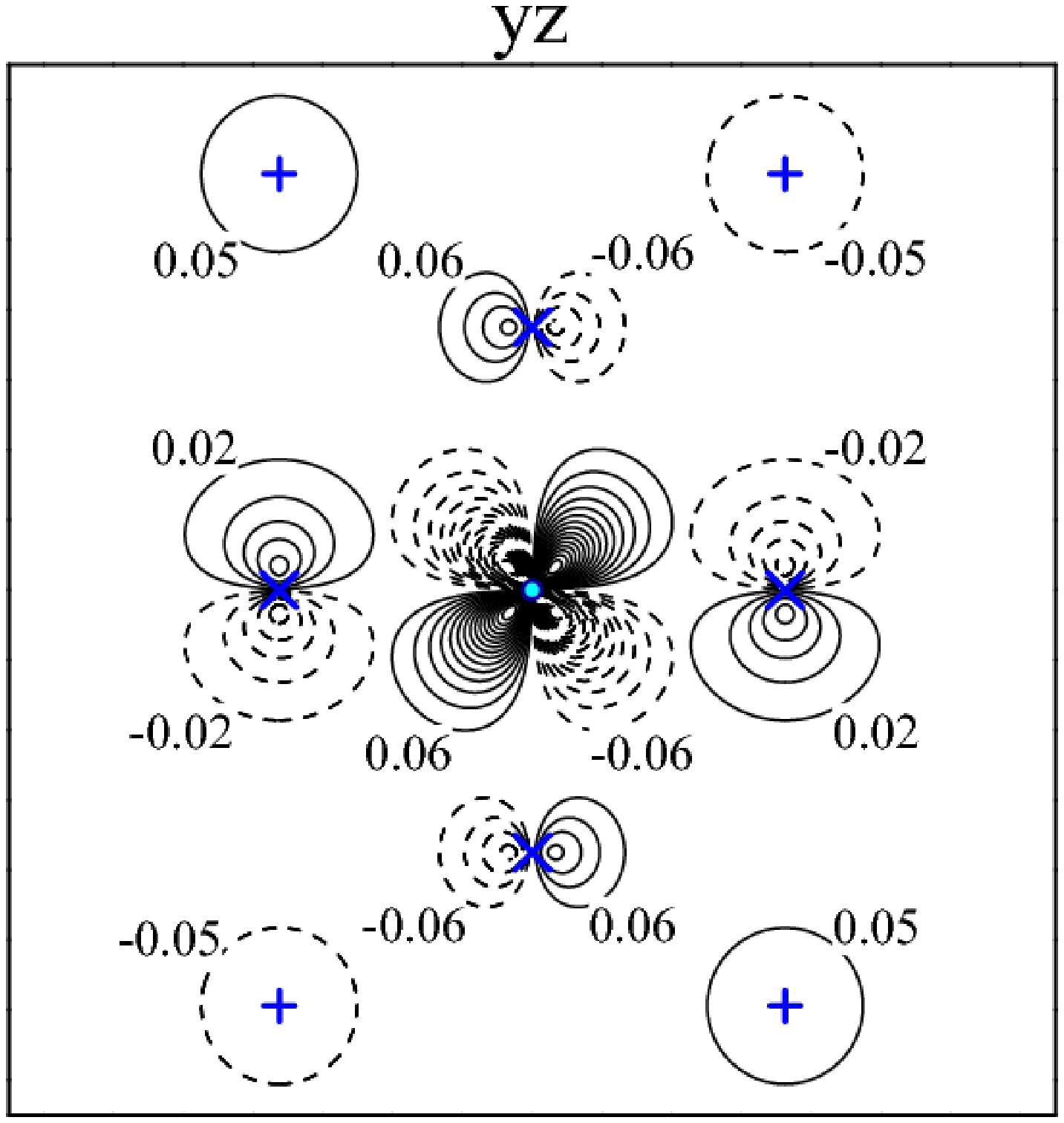}
\end{center}
\caption{(color online) Contour plot of Wannier functions of the $xy$ (left) and $yz$ (right) types
for Sr$_2$VO$_4$. The solid and dashed lines correspond to the positive and negative
values of the Wannier functions. Around each site, the Wannier function
increases/decreases with the step $0.04$ from the values indicated on the graph.
Atomic positions are shown by the following symbols:
$\circ$ (V), $\times$ (O), and $+$ (Em$_1$).}
\label{wannier}
\end{figure}

\begin{figure}[tb]
\begin{center}
\includegraphics[width=6cm]{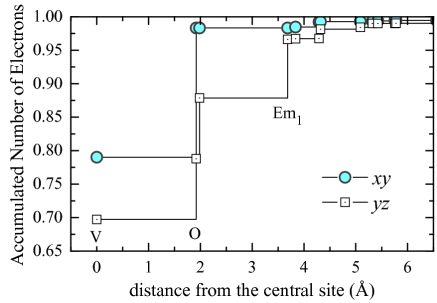}
\end{center}
\caption{Spacial extension of Wannier functions for Sr$_2$VO$_4$: the number of electrons
accumulated around the central V site after adding every new sphere of
neighboring atomic sites.}
\label{charge_wannier}
\end{figure}

  The Wannier functions obtained in such a way are shown in Fig.~\ref{wannier},
and their extension in the real space is illustrated in Fig.~\ref{charge_wannier}.
(color online) Since the $t_{2g}$ band is an antibonding combination of the atomic
V($3d$-$t_{2g}$) and O($2p$) orbitals, the Wannier functions have nodes
located between vanadium and oxygen sites. The $xy$-orbitals appear to be
more localized. This is not surprising because in order to describe the
distribution of the electron density in the $xy$-plane, it is sufficient
to have compact Wannier orbitals, which are confined mainly within
the central V site and the neighboring oxygen sites. Then, the full space within the
plane can be spanned by such Wannier orbitals, periodically repeated in the
$xy$-plane.
On the other hand, the $yz$ and $zx$ orbitals should account for
the penetration of the electron density in rather extended interplane region,
including the areas of weak and nearly constant densities
associated with the empty spaces
of the layered perovskite structure (Em$_1$ in Fig.~\ref{wannier}).
Therefore, the $yz$ and $zx$ orbitals should have a long tails spreading out of
the $xy$-plane.
Another measure of the localization of the Wannier functions is
the expectation values of the square of the position operator
$\langle {\bf r}^2 \rangle$,\cite{MarzariVanderbilt}
which
yields
1.53 and 3.18~\AA$^2$ for the $xy$ and $yz$ ($zx$) orbitals, respectively.

\subsection{Effective Coulomb Interactions}
\label{EffectiveCoulomb}

  The matrix elements of the effective Coulomb interactions in
the $t_{2g}$ band are defined as the energy cost for moving
an electron between two Wannier orbitals, $W_i^m$ and $W_j^{m'}$,
which have been initially populated by $\mathfrak{n}_{im}$ and
$\mathfrak{n}_{jm'}$ electrons:
$$
\Delta_{ij}^{mm'} = E\left[ \mathfrak{n}_{im} + 1, \mathfrak{n}_{jm'} - 1 \right] -
E\left[ \mathfrak{n}_{im} , \mathfrak{n}_{jm'} \right].
$$
For $i$$\ne$$j$, $\Delta_{ij}^{mm'}$ have a meaning of on-site Coulomb interactions
screened by intersite interactions, while for $i$$=$$j$ it stands for the intraatomic
exchange and nonsphericity effects, responsible
for the Hund's first and second rule, respectively.

  These matrix elements can be calculated in two steps,
employing a number of approximations, which have been discussed in details in the
previous paper.~\cite{solo05b}\\
1. First, we perform the standard constraint-LDA (c-LDA) calculations,
where all matrix elements
of hybridization involving the atomic V($3d$)-states
have been artificially switched off.\cite{Gunnarsson} This part takes into
account the screening of on-site Coulomb interactions
caused by the relaxation of the $3d$-atomic basis functions and the
redistribution of the rest of the charge density.
Particularly, the removal of the $3d$ electron
creates at the same site an excessive charge of the $4sp$-type, which compensate the
lack of the $3d$ charge by nearly 50\%.
The values of the on-site Coulomb ($\mathfrak{u}$)
obtained in such an approach vary from 9.7 till 10.4 eV, depending on the
distance between sites $i$ and $j$.
They are considerably smaller than the bare Coulomb interaction between
$3d$ electrons ($\sim$$22$ eV), meaning that the screening of
Coulomb interactions
caused by the relaxation of the basis functions as well as the redistribution
of
non-$3d$ electrons is very efficient and reduces $\mathfrak{u}$ by more than
50\%.
The exchange interaction ($\mathfrak{j}$) is about 1 eV.
Using
only $\mathfrak{u}$ and $\mathfrak{j}$, one can restore the full
$5$$\times$$5$$\times$$5$$\times$$5$ matrix
$\hat{\mathfrak{u}}$$\equiv$$\| \mathfrak{u}_{m m' m'' m'''} \|$ of Coulomb interactions
between atomic $3d$-orbitals, as it is typically done in the
LDA$+$$U$ method.\cite{solo94}\\
2. Then, we switch on the hybridization and evaluate the screening caused by the
change of this hybridization between the
atomic V($3d$)-orbitals and the rest of the basis states
in the random-phase approximation (RPA):
\begin{equation}
\hat{U}(\omega) = \left[ 1 - \hat{\mathfrak{u}}\hat{P}(\omega) \right]^{-1}\hat{\mathfrak{u}}.
\label{eqn:Dyson}
\end{equation}
This scheme implies that different channels of screening can be included
consecutively. Namely, the $\hat{u}$-matrix derived from c-LDA is used as the
bare Coulomb interaction in the Dyson equation (\ref{eqn:Dyson}), and the
$5$$\times$$5$$\times$$5$$\times$$5$
polarization matrix
$\hat{P}$$\equiv$$\| P_{m m' m'' m''' } \|$
describes solely the effects of hybridization
of the V($3d$) states with the O($2p$) and Sr($4d$) states, which leads to
the formation of the distinct V($t_{2g}$) band
in Fig.~\ref{ldados}. The matrix elements of
$\hat{P}$ are given by
\begin{equation}
P_{m m' m'' m'''}(\omega) = 2 \sum_{n {\bf k}} \sum_{n' {\bf k}'}
\frac{(\mathfrak{n}_{n {\bf k}}-\mathfrak{n}_{n' {\bf k}'})
d^\dagger_{m n' {\bf k}'} d_{m' {n \bf k}}
d^\dagger_{m'' n {\bf k}} d_{m''' n' {\bf k}'}}
{\omega - \varepsilon_{n' {\bf k}'} + \varepsilon_{n {\bf k}} +
i\delta (\mathfrak{n}_{n {\bf k}}-\mathfrak{n}_{n' {\bf k}'})},
\label{eqn:Polarization}
\end{equation}
where $\{ \varepsilon_{n {\bf k}} \}$ and $\{ \mathfrak{n}_{n {\bf k}} \}$
are the LDA eigenvalues and occupation
numbers, respectively, for the band $n$ and momentum ${\bf k}$
in the first Brillouin zone,
and $d_{m n {\bf k}}$$=$$\langle \chi_m | \psi_{n {\bf k}} \rangle$
is the projection of the LDA eigenstate $\psi_{n {\bf k}}$ onto
the atomic $3d$ orbital $\chi_m$.

\begin{figure}[h!]
\begin{center}
\resizebox{8cm}{!}{\includegraphics{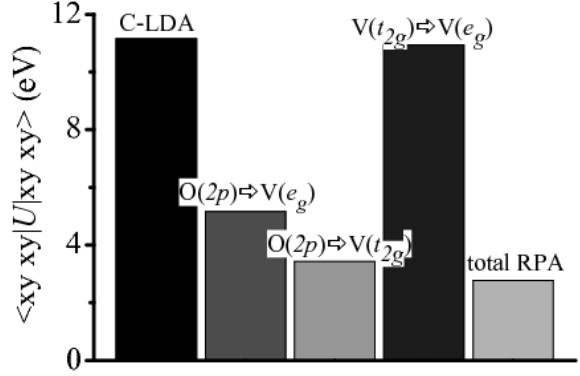}}
\end{center}
\caption{\label{fig.Upartial}
Screened intra-orbital Coulomb interaction
in constraint-LDA and RPA.
The partial contributions show the screening caused by
separate interband transitions
in the
polarization function (\protect\ref{eqn:Polarization}).}
\end{figure}

  Results of these calculations are shown in Fig.~\ref{fig.Upartial}.
In this case one can easily separate different contributions to the RPA screening caused
by the interband O($2p$)$\rightarrow$V($e_g$), O($2p$)$\rightarrow$V($t_{2g}$),
and V($t_{2g}$)$\rightarrow$V($e_g$) transitions in the polarization function
(\ref{eqn:Polarization}).
Note that the screening caused by the V($t_{2g}$)$\rightarrow$V($t_{2g}$) transitions
in the polarization function is not included here because all Coulomb interactions
within $t_{2g}$ band will be treated rigorously using the PIRG method, beyond RPA.
Since the V-states of the $t_{2g}$ and $e_g$ symmetry are practically not mixed
by transfer interactions in the two-dimensional perovskite lattice, the screening of
on-site Coulomb interactions associated with the V($t_{2g}$)$\rightarrow$V($e_g$)
transitions appears to be weak.
On the other hand, the screening caused by the
O($2p$)$\rightarrow$V($e_g$) and O($2p$)$\rightarrow$V($t_{2g}$) transitions
is very efficient, so that the characteristic parameters of inter-orbital Coulomb
interactions are strongly reduced in comparison with c-LDA, typically
till $2.6$-$2.8$ eV.

\begin{figure}[h!]
\begin{center}
\resizebox{8cm}{!}{\includegraphics{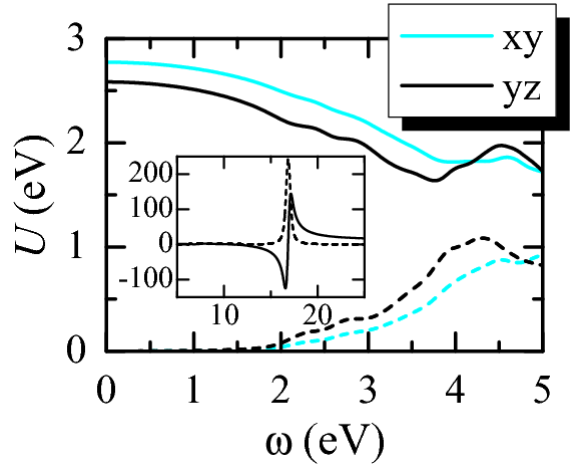}}
\end{center}
\caption{\label{fig.UERPA}
(color online) Frequency-dependence of
intra-orbital Coulomb interaction in RPA.
The solid (dashed) lines represent real (imaginary) parts.
The inset shows high-frequency part of $U(\omega)$.}
\end{figure}

  This, however, implies a strong $\omega$-dependence of the screened Coulomb interaction,
as it immediately follows from the Kramers-Kronig transformation in RPA:\cite{FerdiGunnarsson}
$$
{\rm Re}\hat{U}(\omega) = \hat{\mathfrak{u}} - \frac{2}{\pi}
{\cal P} \int_0^\infty d \omega' \frac{\omega' | {\rm Im} \hat{U}(\omega') |}
{\omega^2-{\omega'}^2}.
$$
Indeed, the change of ${\rm Re}\hat{U}(\omega)$ at $\omega$$=$$0$ is directly related with
the spectral weight of $|{\rm Im} \hat{U}(\omega)|$ at finite $\omega$
(Fig.~\ref{fig.UERPA}).
This spectral weight can contribute to the renormalization
of the low-energy part of the spectrum through the self-energy effect.
The latter can be evaluated in the GW approach,\cite{FerdiGunnarsson,Hedin,Hanke}
where the self-energy is given by the convolution of $\hat{U}(\omega)$ with the
one-particle Green function $\hat{G}(\omega)$ for the $t_{2g}$ band:
$$
\hat{\Sigma}(\omega)=\frac{i}{2 \pi} \int d\omega'
\hat{G}(\omega + \omega') \hat{U}(\omega').
$$
Then, the renormalization factor is
$$
Z = \left[ \left . 1 - \partial {\rm Re} \Sigma / d \omega \right|_{\omega=0} \right]^{-1},
$$
and the numerical estimates yield $Z$$\approx$$0.8$.  The momentum dependence ot $Z$ is negligibly small.
The imaginary part of $\Sigma$ in the low-frequency region
is also small and can be neglected.

  A new aspect of the RPA screening in the tetragonal compound is the anisotropy of screened Coulomb
interaction. Since the ($xy$) and ($yz$,$zx$) $t_{2g}$ levels belong to different representations
of the point symmetry group, the corresponding matrix elements of the screened Coulomb
interaction $\hat{U}$ can be also different. For the static interactions this
anisotropy is of the order of $0.2$ eV, where the matrix elements between
more extended $yz$ and $zx$ orbitals
are better screened.

The intersite Coulomb interaction can be also derived and
the effective Hamiltonian with the intersite Coulomb term may also be solved by our low-energy solver.
However, in this paper, we do not consider this effect because the intersite Coulomb interaction is substantially smaller.
For example, the value of intersite Coulomb interactions $V$ derived from the standard c-LDA approach
(part 1 of the above procedure) for the cubic
SrVO$_3$ is about $1.2$ eV, which is not negligible.
However, by taking into account the relaxation effects associated with the
V($t_{2g}$)$\rightarrow$V($e_g$) transition, which can be also treated in the framework of c-LDA,
the intersite Coulomb interaction
$V$ is reduced drastically from $1.2$ till $0.3$ eV.\cite{solo05b}
Furthermore,
we expect this value to be further reduced by taking into account the remaining
O($2p$)$\rightarrow$V($t_{2g}$)
and V($t_{2g}$)$\rightarrow$V($e_g$)
transitions in the polarization functions.

The effective low-energy Hamiltonian is now derived from the Coulomb interaction obtained in \S \ref{EffectiveCoulomb} and the transfer.  The
transfer is given by the downfolded tight binding parameter $h_{i,j}^{m,m'}$ remormalized by $Z$, where the self-energy effect reduces the band width. Namely, we take the transfer $t_{i,j}^{m,m'} =Zh_{i,j}^{m,m'}$.

\section{Low-Energy Effective Model}

The effective low-energy Hamiltonian for the $t_{2g}$ orbital obtained by the downfolding described in \S \ref{High-Energy} has a form of the extended Hubbard model and is given by
\begin{eqnarray}
H_{\rm eff}&=&H_{\rm kin}+H_{\rm int}
\label{eqn_effham}
\\
H_{\rm kin}&=&\sum_{<i,j>}\sum_{m,m',\sigma}
t^{mm'}_{ij}c^{\dag}_{im\sigma}c_{jm'\sigma} \\
H_{\rm int}&=&\sum_{i,m}U^{m}n_{im\uparrow}n_{im\downarrow} \nonumber \\
&+&\sum_{i,\sigma,\sigma'}\sum_{m < m'}
{K}^{m,m'}_{\sigma,\sigma'}n_{im\sigma}n_{im'\sigma'}\nonumber \\
&-&\sum_{i,m,m'}J^{m,m'}
(c^{\dag}_{im\uparrow}c_{im\downarrow}
 c^{\dag}_{im'\downarrow}c_{im'\uparrow}\nonumber \\
&+&c^{\dag}_{im\uparrow}c_{im'\downarrow}
 c^{\dag}_{im\downarrow}c_{im'\uparrow}),\\
{K}^{m,m'}_{\sigma,\sigma'}&=&
\left\{
\begin{array}{cc}
K^{m,m'}-J^{m,m'}&(\sigma=\sigma')\\
K^{m,m'}&(\sigma\ne\sigma'),\\
\end{array}
\right.
\end{eqnarray}
where $c^{\dag}_{im\sigma}(c_{im\sigma})$ is the creation (annihilation) operator of a conduction electron with spin $\sigma =(\uparrow, \downarrow)$, and orbital $m$=($xy,yz,zx$: $t_{2g}$ orbitals) at site $i$. $t^{mm'}_{ij}$ is the hopping matrix and $n_{im\sigma}=c^{\dag}_{im\sigma}c_{im\sigma}$. $U^{m}({K}^{m,m'})$ is the intra-orbital (inter-orbital) Coulomb interaction and $J^{m,m'}$ is the exchange interaction, which gives rise to the Hund's rule coupling and pair hopping.

For ${\rm Sr_{2}VO_{4}}$, estimate of each parameter obtained in the procedure explained in \S \ref{High-Energy} are given as  $U^{xy}=2.77$ eV,
$U^{yz}=U^{zx}=2.58$ eV, $K^{xy,yz}=K^{zx,xy}=1.35$ eV, $K^{yz,zx}=1.28$ eV, $J^{xy,yz}=J^{zx,xy}=0.65$ eV, and $J^{yz,zx}=0.64$ eV.
These values approximately hold the relation $U \sim K+2J$, which shows that the effective low-energy Hamiltonian satisfies rotational
invariance in spin and orbital spaces.

\begin{figure}[tb]
\begin{center}
\includegraphics[width=8cm]{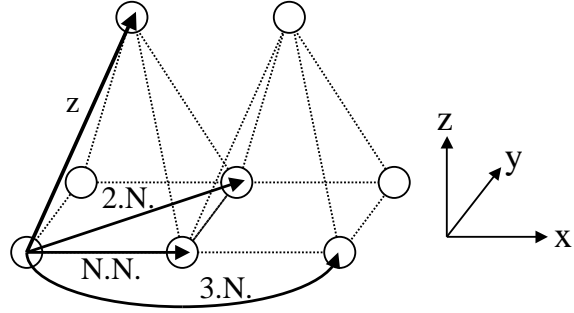}
\end{center}
\caption{Transfer directions of 3$d$ $t_{2g}$ electrons on V sites in ${\rm Sr_{2}VO_{4}}$; Open circles denote the lattice sites
of V. Nearest neighbor sites (N.N), 2nd neighbor sites (2.N.), 3rd neighbor sites (3.N.), and $z$ direction (z) transfers are depicted.}
\label{tra_dir}
\end{figure}

The transfer direction is shown in Fig. \ref{tra_dir}. By considering transfers until third neighbor sites, tight-binding approximation
can reproduce the LDA band structure (Fig. \ref{band}).
Off-diagonal hoppings for orbital indices become zero or quite small, so that
we show the diagonal hopping amplitude in Table \ref{hoppings}. However, note that all components of hopping matrices within third neighbor sites are considered in our calculations.

\begin{table}
\caption{Diagonal components of hopping matrices for orbital indices, where the energy is in eV.
}
\begin{tabular}{ccccc} \hline
  & N.N. & 2.N.& 3.N. & z \\ \hline
$t^{xy,xy}$ & -0.218 & -0.075 & 0.000 &  0.000 \\ 
$t^{yz,yz}$ & -0.045 & -0.010 & 0.000 & -0.016 \\ 
$t^{zx,zx}$ & -0.194 & -0.010 & 0.018 & -0.016 \\ \hline
\end{tabular}
\label{hoppings}
\end{table}

Because the 2D character of ${\rm Sr_{2}VO_{4}}$ is strong, in the present study, we consider 2D $N=L\times L$ lattice first and then
 $N=L\times L\times L_{z}$ lattices to understand three-dimensional effects.

\section{Conventional Approach}

Before going into detailed results obtained from the PIRG method, we first discuss results obtained from more conventional approach. 
The density of states (DOS) of ${\rm Sr_{2}VO_{4}}$ calculated by the LDA is shown in Fig. \ref{ldados}. 
A sharp peak around the Fermi level appears in the DOS, which indicates that the ground state is a paramagnetic metal. 
A similar result has been obtained by Singh. {\it et al}. \cite{sing91}. 
These results show that the LDA result contradicts the experimental results for ${\rm Sr_{2}VO_{4}}$. 
On the other hand, as we will see later, the Hartree-Fock approximation for the effective Hamiltonian, also underlying the commonly used LDA+U approach,\cite{solo94} indicates that the ground state is a complete ferromagnetic insulator. These are again not supported by the experimental results. 
Since LDA and HFA offer completely different phases, it is desired to apply a more reliable method suited for strongly correlated electron systems.

It is widely accepted that LDA is insufficient for strongly correlated electron systems because of neglecting spatial as well as dynamical fluctuations arising from strong correlation and quantum effects. In order to go beyond LDA, so-called LDA+U\cite{anis91,solo94,anis97a} and LDA+DMFT\cite{anis97b,lich98,held01,held02} methods were proposed. In LDA+U, the Coulomb interactions are treated within HFA and the ground state is obtained within the framework of single Slater determinant, so that it is also difficult to take into account effects of electron correlations and quantum fluctuations. On the other hand, in the LDA+DMFT which combines LDA calculation with the dynamical mean field theory (DMFT) \cite{kura85,prus95,geor96}, it is possible to calculate dynamical properties such as spectral functions and the optical conductivity with dynamical fluctuations being taken into account. 
However, within DMFT, it is not easy to consider spatial fluctuations as well as possible nontrivial spatial order, because the Hubbard type lattice model is reduced to an effective impurity problem in DMFT with spatial fluctuations being ignored and its extension to a large unit cell is not easy either.

\section{Path-integral renormalization group method}
Recently, we proposed a new algorithm, the LDA combined with the PIRG method\cite{imai05}. 
In order to solve the effective Hamiltonian eq. (\ref{eqn_effham}), we employ PIRG method~\cite{imad00,kash01,mori02,wata03,mizu04}. 
The PIRG can systematically include symmetry breakings and spatial fluctuations neglected in DMFT.
PIRG can be applied to various systems which can not be treated by other numerical techniques such as the quantum Monte Carlo method
because the sign problem does not appear explicitly. Furthermore a Slater determinant, given as the eigenvector of the HFA, is usually employed as the
initial state of PIRG. Since quantum fluctuations are evaluated by a systematic expansion of the basis states in the renormalization
process of PIRG, we can discuss correlation effects in comparison with results of the HFA in a systematic renormalization procedure.
Therefore this new scheme enables us to investigate the ground state properties for strongly correlated electron systems, and is suitable
method for evaluating physical properties from the first-principles.

We briefly review PIRG method in this subsection.
According to the path-integral formalism, the ground state wave function $|\psi_{g}\rangle$ can be obtained from
\begin{eqnarray}
|\psi_{g}\rangle =\lim_{\tau \rightarrow \infty}
\exp{[-\tau H]|\phi_{0}\rangle},
\end{eqnarray}
where $|\phi_{0}\rangle$ represents the initial state. This projection process is performed in the imaginary time direction $\tau$. In PIRG, the ground state is approximately represented by a linear combination of proper basis states,
\begin{eqnarray}
|\psi_{g}\rangle \approx \sum^{L}_{i=1}\omega_{i}|\phi_{i}\rangle.
\end{eqnarray}
Since basis states should be chosen to give the lowest energy within given dimensions of the Hilbert space, we have to optimize the coefficient $\omega_{i}$ and relevant basis states $|\phi_{i}\rangle$, which constitute the process of our numerical renormalization. This procedure is symbolically written as
\begin{eqnarray}
|\phi_{i}\rangle = \exp{({-\Delta \tau H})}|\phi^{(0)}_{i}\rangle.
\label{eqn_prj}
\end{eqnarray}
The optimization of $|\phi_{i}\rangle$ and $\omega_{i}$ is performed in the following way:
We employ Slater determinants as the basis functions.
Candidates of the Slater-determinant basis states $|\phi_{i}\rangle$ are generated from the operation of $\exp{(-\Delta \tau H)}$ with a
small $\Delta \tau$ to the previous basis  $|\phi^{(0)}_{i}\rangle$. To operate $\exp[-\Delta\tau H]$,
the decomposition $\exp{[-\Delta \tau H]} \simeq \exp{[-\Delta \tau H_{\rm kin}]}\exp{[-\Delta \tau H_{\rm int}]}$ is used and
the operation of $\exp{[-\Delta \tau H_{\rm int}]}$ is taken by introducing a Stratonovich-Hubbard transformation~\cite{imad00}.
This transformation generates more than one candidate for the next generation basis.
Among the candidates, we employ the optimized basis which gives the lowest energy eigenvalue in the fixed subspace and discard other candidates.
The eigenvalue is computed from the ground state of the generalized eigenvalue problem:
\begin{eqnarray}
E=\frac{\sum^{L}_{i,j}\left[H\right]_{ij}\omega_{i}\omega_{j}}
{\sum^{L}_{i,j}\left[F\right]_{ij}\omega_{i}\omega_{j}},
\end{eqnarray}
where
\begin{eqnarray}
\left[H\right]_{ij}&=&\langle \phi_{i}|H|\phi_{j}\rangle,\\
\left[F\right]_{ij}&=&\langle \phi_{i}|\phi_{j}\rangle.
\label{HF}
\end{eqnarray}
We repeat the operation of $\exp[-\Delta\tau H]$ and the truncation by selecting the optimized basis within a fixed dimension $L$.  The former
process expands the number of basis functions beyond $L$ and the latter truncation reduces the number to $L$ again. After sufficient number of
these iterations, the energy and the basis functions converge within the fixed $L$.  The fixed point of the renormalization process is then completed at the fixed $L$.

The ground state obtained in this restricted Hilbert space is an approximate ground state of the full Hilbert space.
Then we further gradually increase $L$ and repeat the iteration to obtain the fixed point at each $L$.
To estimate the correct ground state properties, one needs extrapolations of the number of states $L$ to the dimension of the full
Hilbert space.  According to the general extrapolation procedure, the following relation
\begin{eqnarray}
\langle H \rangle - \langle H \rangle_{g} \propto \Delta E
\label{Evariance}
\end{eqnarray}
is satisfied, where $\langle H \rangle_{g}$ is the energy expectation value for the true ground state and $\Delta E$ is the energy variance defined by,
\begin{eqnarray}
\Delta E =\frac{\langle H^{2} \rangle - \langle H \rangle^{2}}
{\langle H \rangle^{2}}.
\end{eqnarray}
For an arbitary physical quantity $A$, similar relations hold as
\begin{eqnarray}
\langle A \rangle - \langle A \rangle_{g} \propto \Delta E.
\label{Avariance}
\end{eqnarray}
In this paper, we examine dependences of these physical quantities on the energy variances in detail. We increase $L$ until the dependence converges reliably to the linear behavior and the linear fitting is taken in this large $L$ region.

We take this extrapolation as far as the linearities as eq. (\ref{Evariance}) and (\ref{Avariance}) are satisfied in the large $L$ region.   Following the above procedures, we can estimate the energy eigenvalue and other physical quantities in the ground state with a systematic extrapolation to the full Hilbert space.

Recently, the quantum-number projection method was implemented in PIRG~\cite{mizu04}, which restores the symmetry of the Hamiltonian in the renormalization process and improves the accuracy.
For example, the extrapolated ground state energy is $-66.879$, which is within the statistical error of the result of quantum Monte Carlo calculation ($E_{g}=-66.866 \pm 0.050$) for the single orbital Hubbard model at half filling on a $6\times6$ square lattice with the nearest neighbor hopping $t=1$ at the onsite Coulomb interaction $U=4$.

In this process, by its construction, there is no negative sign problem in any models, which appears in the quantum Monte Carlo method. PIRG enables us to accurately study physical properties in various lattices and at arbitrary fillings. Therefore this method is suitable for the present purpose of investigating properties of an extended Hubbard-type models in comparison with existing numerical algorithms.

Here note that in order to apply PIRG to the multi-orbital Hubbard model, we have to treat the projection processes of intra-, inter-orbital interactions, and the Hund's rule coupling. However, in order to reduce the computational time, we skip projection processes of applying the inter-orbital and the Hund's rule coupling terms in the Hamiltonian of $\exp [-\tau H]$. This saving of the computational time does not seem to damage the efficiency of the projection process.  Of course, the lowering of the energy evaluated in the generalized eigenvalue problem is evaluated by the correct form of the Hamiltonian including the exchange terms.

Therefore, the lowest energy state  is obtained if the renormalization projection procedure eq. (\ref{eqn_prj}) is taken until
the convergence. In this text, we also note that the projection eq. (\ref{eqn_prj}) generates a local minimum of the Hamiltonian
$H$ within the fixed number of basis. It may be necessary to seek for the global minimum by taking as many as possible candidates of
the initial states $|\phi_{0} \rangle $. The real ground state may be obtained from the comparison of these local minima.
This is particularly important if several phases with different symmetry are competing as in the present case.

\section{Results}
In this section, we show physical quantities for the effective low-energy Hamiltonian of  ${\rm Sr_{2}VO_{4}}$ and discuss ground state properties. We first apply the usual Hartree-Fock approximation to the effective Hamiltonian eq.(\ref{eqn_effham}). In the next subsection, we show results obtained by the PIRG and discuss the relevance to experimental results.

 Here, in order to study correlation effects systematically, we introduce the interaction parameter $\lambda$, which scales the amplitudes of all the matrix elements ($U, U', J$) with the ratio of those interactions being fixed. The realistic value in the downfolding Hamiltonian corresponds to $\lambda=1.0$. Namely, we study the low-energy Hamiltonian by extending eq. (\ref{eqn_effham}) as
\begin{eqnarray}
H_{\rm eff}=H_{\rm kin}+\lambda H_{\rm int}
\end{eqnarray}
to examine the correlation effects clearly.

Since ${\rm Sr_{2}VO_{4}}$ has a layered perovskite structure with strong two-dimensional anisotropy, we first consider two-dimensional square lattices. Three-dimensional effects are studied in the later part of this paper.

\subsection{Hartree-Fock results}
The Slater determinants are usually employed in the HFA and for the basis states in PIRG method as well. Crucial difference of PIRG from
HFA is that PIRG allows representing an approximate ground state by increasing number of Slater determinants while HFA represents
the ground state only by a single Slater determinant.  In other words, PIRG is able to include quantum fluctuations by improving
HFA results, so that we first study this Hamiltonian within HFA as a starting point.

Here we show results obtained by the HFA for the two-dimensional square lattices. Since the amplitude of orbital off-diagonal hoppings is
much smaller than that of diagonal hoppings, each base ($|xy\rangle$, $|yz\rangle$, and $|zx\rangle$) is nearly orthogonal.
\begin{figure}[b]
\begin{center}
\vspace{-5mm}
\includegraphics[width=8cm]{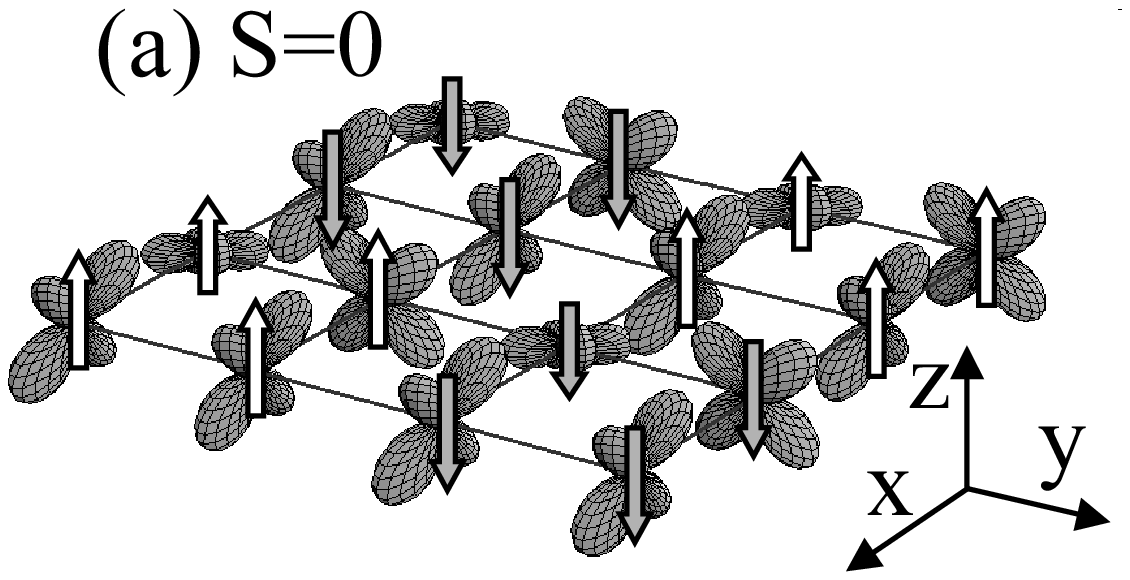}
\vspace{-5mm}
\includegraphics[width=8cm]{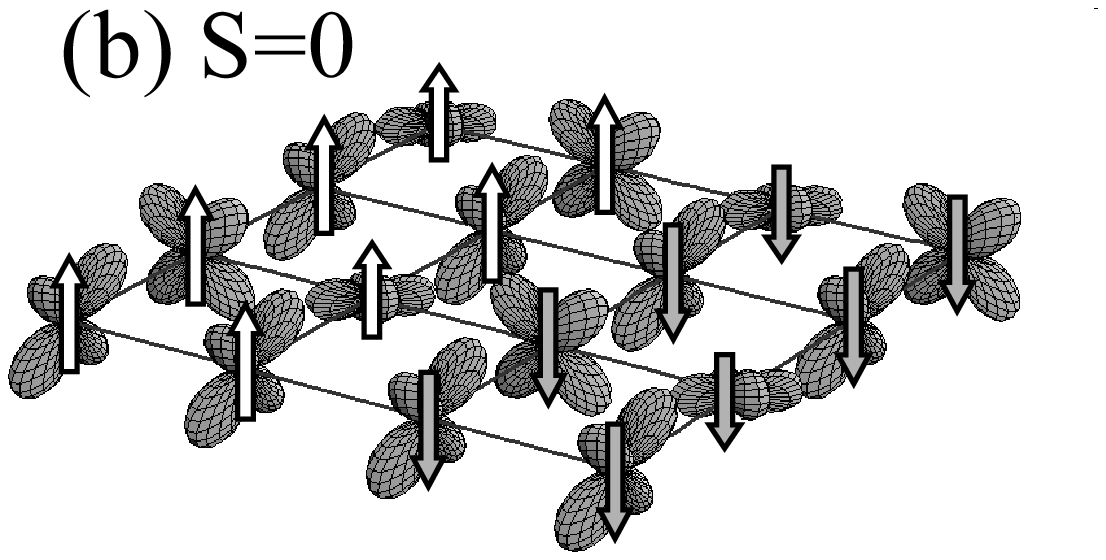}
\vspace{-5mm}
\includegraphics[width=8cm]{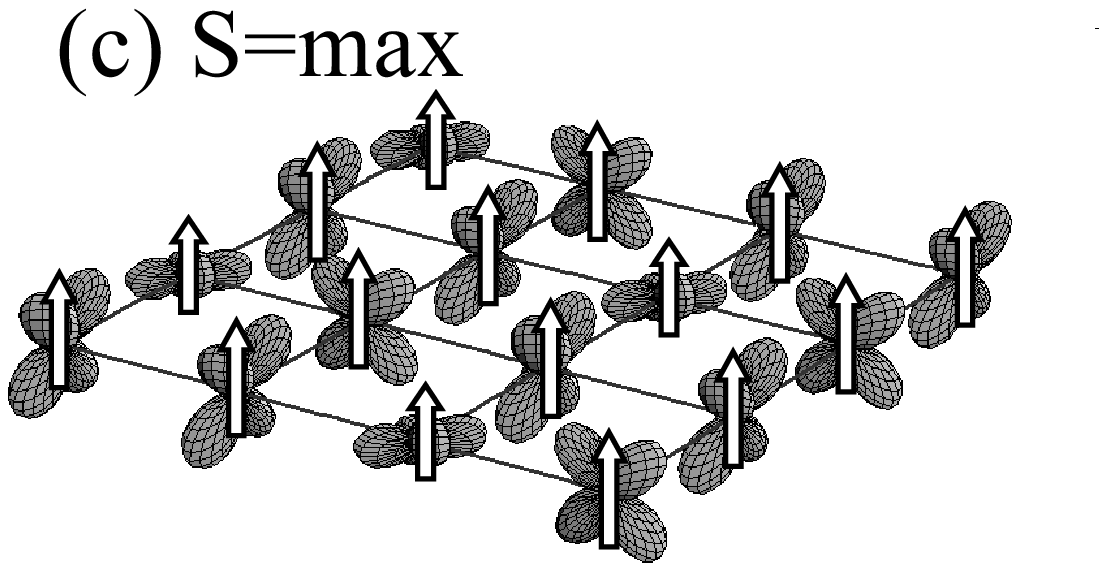}
\vspace{-5mm}
\includegraphics[width=8cm]{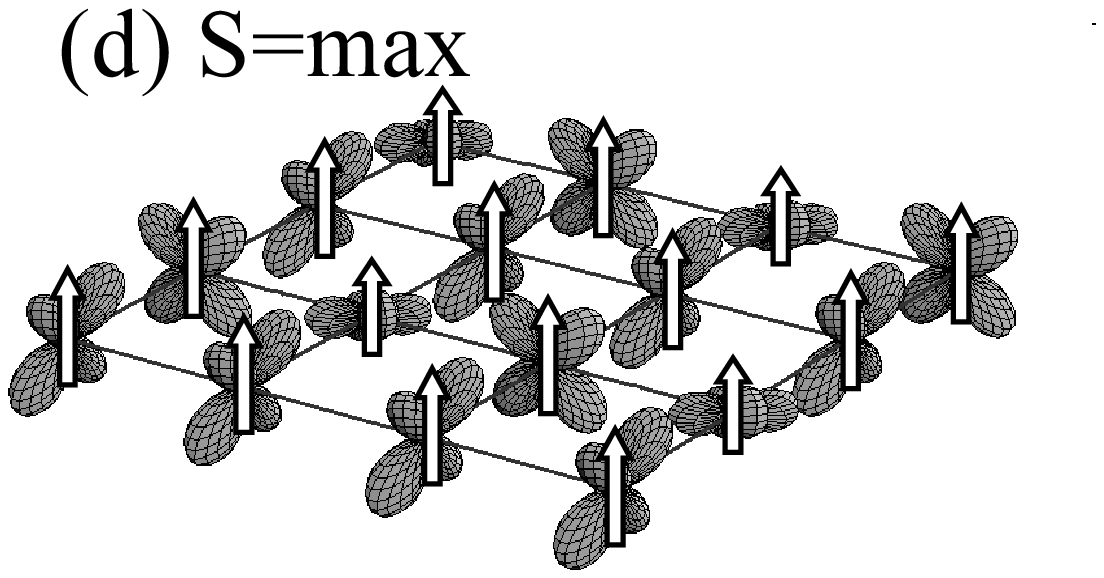}
\end{center}
\caption{Possible spin and orbital patterns obtained as metastable states of HFA; the $S$=0 state ((a) and (b)) and the complete
ferromagnetic state ((c) and (d)) when $\lambda \sim 1.0$. Up (down) arrow represents the up (down) spin. The dominant orbital
and spin patterns are illustrated only schematically and the actual solution contains some linear combination of other orbitals.}
\label{order}
\end{figure}
In Fig. \ref{order}, we show several possible spin and orbital patterns in the $S$=0 and complete ferromagnetic ($S$=max) sectors
in the $x$-$y$ plane, respectively.
For the Fig. \ref{order} (a) and (b) structures, typical order parameters are given by;
$
o^{o}=\frac{1}{2}\sum_{\sigma}
|\langle n^{o}_{xy \sigma} \rangle
-\langle n^{o}_{yz \sigma} \rangle |
$
and
$
m^{o}_{\alpha}=
\frac{1}{2}|\langle n^{o}_{\alpha \uparrow} \rangle
-\langle n^{o}_{\alpha \downarrow} \rangle |
$
 at odd-numbered sites to the $x$ direction, where $|\langle n^{o}_{\alpha \sigma} \rangle |$ represents the expectation value of electron
number with spin $\sigma$ and orbital $\alpha$=$xy, yz$. Similarly,
$
m^{e}_{zx}=
\frac{1}{2}|\langle n^{e}_{zx \uparrow} \rangle
-\langle n^{e}_{zx \downarrow} \rangle |,
$
at even-numbered sites to the $x$ direction.
For the Fig. \ref{order} (c) structures,
$
o^{o}=
\frac{1}{2}|\langle n^{o}_{xy \uparrow} \rangle
-\langle n^{o}_{zx \uparrow} \rangle |
$
and
$
o^{e}=
\frac{1}{2}|\langle n^{e}_{yz \uparrow} \rangle
-\langle n^{e}_{zx \uparrow} \rangle |
$
at odd and even sites to the $y$ direction are the relevant order parameters.
For the Fig. \ref{order} (d) structures, we take
$
o^{o}=
\frac{1}{2}|\langle n^{o}_{xy \uparrow} \rangle
-\langle n^{o}_{yz \uparrow} \rangle |
$
at odd sites to the $x$ direction.

These are obtained at least as metastable states of the HF solution. Because many of dominant orbitals are different from those of the nearest neighbor sites and rather orthogonal in all the panels of Fig. \ref{order}, the nearest-neighbor exchange coupling has primarily ferromagnetic sign because of the Goodenough-Kanamori rule.

However, this ferromagnetic exchange is not absolutely dominant because the orbitals are nonnegligibly mixed and the nearest-neighbor orbitals are not strictly orthogonal.  Furthermore, substantial third neighbor hoppings between the same orbitals exist and causes the antiferromagnetic coupling between the third-neighbor sites.
Due to the antiferromagnetic interaction between the third neighbor sites, frustration is induced and a complex ordering of Fig. \ref{order} (a) appears. Fig. \ref{order} (b) is also a possible order where the spin stripe structure appears to the $x$ direction.

\begin{figure}[tb]
\begin{center}
\includegraphics[width=8cm]{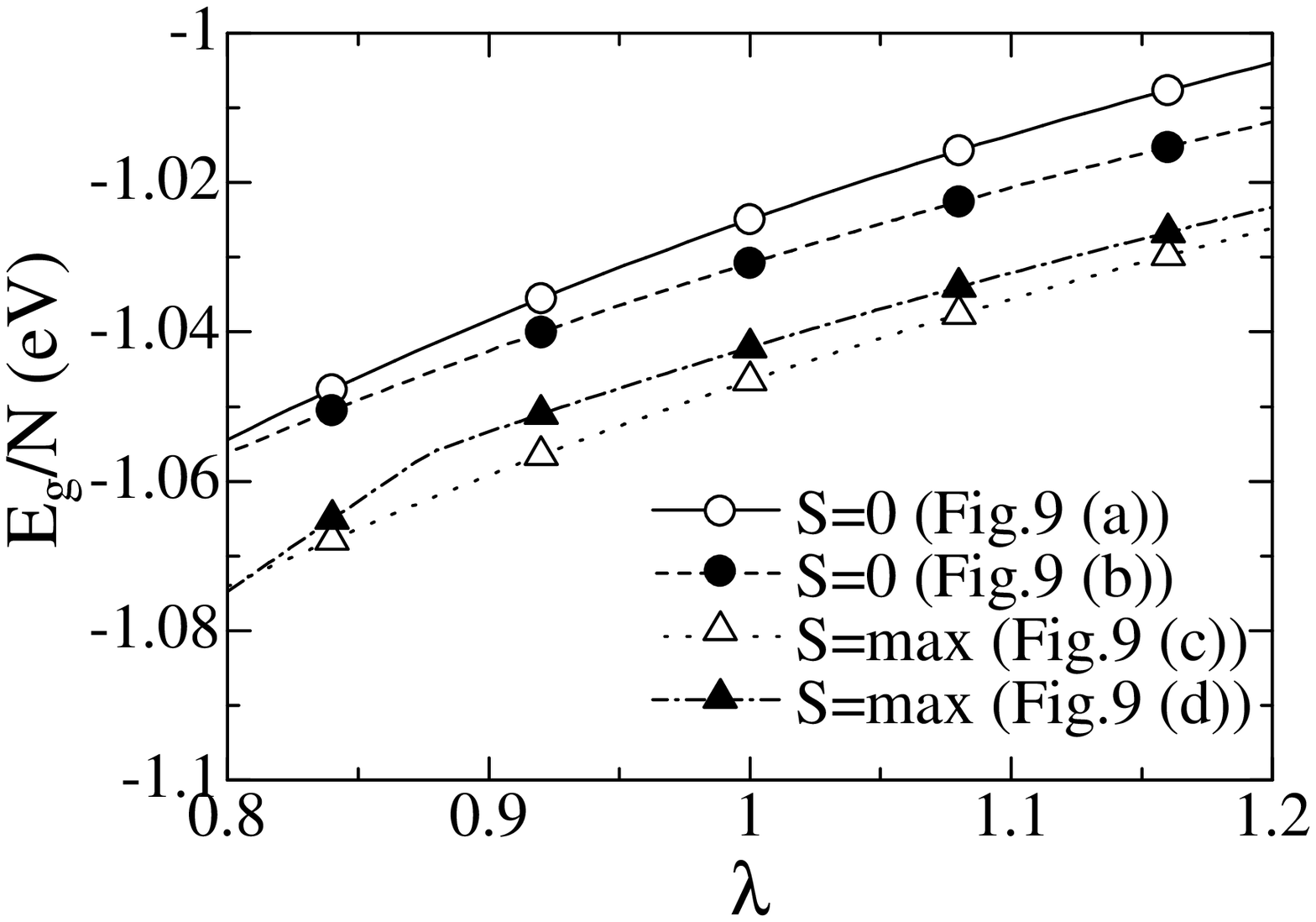}
\end{center}
\caption{Total energy per site as a function of the interaction parameter $\lambda$ obtained by HFA for each spin and orbital pattern in Fig. \ref{order} for the two-dimensional square lattices.
}
\label{eg_hf}
\end{figure}

Figure \ref{eg_hf} shows the total energies per site in the interval $0.8 < \lambda < 1.2$.
Although the energy for the order (b) is lower than that of (a) within HFA in $S$=0 sector at $\lambda \sim 1.0$, the difference of each energy is very small ($\sim$ 0.01eV).
On the other hand (c) is the lowest-energy state for $S$=max and becomes the absolute ground state within the Hartree-Fock approximation.
However, the energy of (d) state is close to that of (c), where the difference of the energy per site between (c) and (d) is about 0.005eV at $\lambda \sim 1.0$.
Because of the ferromagnetic and antiferromagnetic interactions coexisting in this system, the ordering structure of the spin and orbital may become complicated at the realistic parameter value of the material ($\lambda \sim 1.0$). Therefore we can not exclude the possibility of other more complicated ordered state. Although it is difficult to determine the Hartree-Fock ground state, the unit cell of ${\rm Sr_{2}VO_{4}}$ becomes large and nontrivial in any case.

By including quantum fluctuations by PIRG, the order (a) becomes ground state and the order (c) becomes most stable in $S$=max sector in our calculations, which will be discussed in the next subsection. Therefore, hereafter we mainly consider the order (a) and (c) in the insulating phases in this subsection.

\begin{figure}[tb]
\begin{center}
\includegraphics[width=8cm]{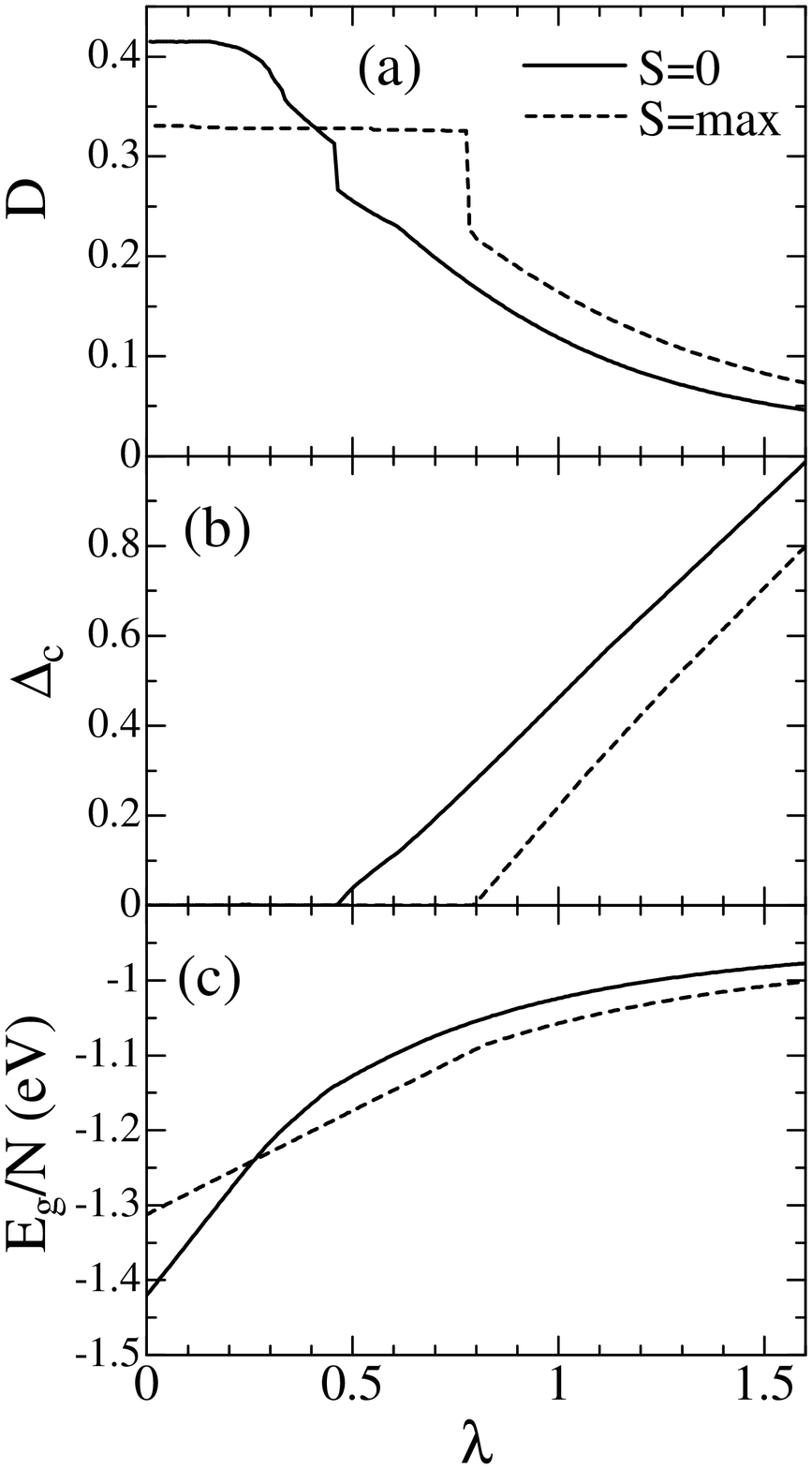}
\end{center}
\caption{(a) Double occupancy $D$, (b) charge gap $\Delta_{c}$, and (c) total energy per site as a function of the interaction parameter $\lambda$ obtained by HFA for the two-dimensional square lattices.
The solid line represents $S$=0 state in Fig. \ref{order} (a) and the dashed line is $S$=max state in Fig. \ref{order} (c).
Note that the configuration Fig. \ref{order} (a) gives the lowest energy among $S$=0 states.
}
\label{hfrslt}
\end{figure}
In Fig. \ref{hfrslt}, we show the double occupancy, charge gap, and total energy per site in total $S$=0 sector (Fig.\ref{order}(a)) and the complete ferromagnetic ($S$=max) state (Fig.\ref{order}(c)). The double occupancy $D$ and the charge gap $\Delta_{c}$ are defined as
\begin{eqnarray}
D&=&\frac{1}{N}\sum_{i}\sum_{(m,\sigma)<(m',\sigma')}
\langle n_{im\sigma}n_{im'\sigma'}\rangle ,\\
\Delta_{c}&=&\epsilon^{us}_{\bf k}-\epsilon^{os}_{\bf k}.
\end{eqnarray}
The charge gap is defined as the difference between the energies of the lowest unoccupied state $\epsilon^{us}_{\bf k}$ and the highest occupied state $\epsilon^{os}_{\bf k}$, where $\epsilon_{\bf k}$ represents the dispersion of electrons

When $\lambda=0$, $D\sim 0.42$ at $S$=0 while $D\sim 0.33$ is obtained at $S$=max.
With increasing interaction parameter $\lambda$, the jumps of double occupancies appear around $\lambda \sim 0.45$ ($S$=0) and $\lambda \sim 0.80$ ($S$=max) and accordingly each charge gap begins to open continuously.
These results indicate that the 1st order metal-insulator transition occurs at $\lambda \sim 0.45$ ($S$=0) and $\lambda \sim 0.80$ ($S$=max) within the HFA.

Within HFA, the complete ferromagnetic and insulating state becomes always more stable than the $S$=0 state when $\lambda >0.25$. Therefore when $\lambda=1.0$, which is the realistic value estimated by the LDA, the HFA predicts the complete ferromagnetic and insulating ground state. This result does not seem to be consistent with the experimental results. However, the difference between the total energy of $S$=0 and that of $S$=max is the order of $10^{-2}$ eV because of the competing ferromagnetic and antiferromagnetic interactions. This system appears to be very sensitive to the control of electron correlations. To obtain ground state properties, we have to consider effects of quantum fluctuations, which are not included within HFA.

Next, let us consider the magnetic and orbital orders. Equal-time structure factors are defined as
%
\begin{eqnarray}
S(\mib q)&=&\frac{1}{N}\sum_{i,j}
 \langle \mib S_{i} \cdot \mib S_{j} \rangle
 \exp({{\rm i}\,{\mib q} \cdot (\mib r_{i}-\mib r_{j}})),
\label{eqn_cf_spin}
 \\
T(\mib q)&=&\frac{1}{N}\sum_{i,j}
 \langle \mib T_{i} \cdot \mib T_{j} \rangle
 \exp({{\rm i}\,{\mib q} \cdot (\mib r_{i}-\mib r_{j}})),
\label{eqn_cf_orbital}
\end{eqnarray}
where $\mib S_{i}$ and $\mib T_{i}$ are the spin and orbital operators as follows,
%
\begin{eqnarray}
\mib S_{i}&=&\frac{1}{2}\sum_{m,\sigma,\sigma'}
c^{\dag}_{im\sigma}{\mib \sigma}_{\sigma,\sigma'}c_{im\sigma'},
 \\
\mib T_{i}&=&\frac{1}{\sqrt{N}}\sum_{m,m',\sigma}
c^{\dag}_{im\sigma}{\mib \lambda}_{m,m'}c_{im'\sigma},
\end{eqnarray}
where ${\mib \sigma}$ and ${\mib \lambda}$ are the Pauli and the Gell-Mann matrices, respectively.

\begin{figure}[tb]
\begin{center}
\includegraphics[width=8cm]{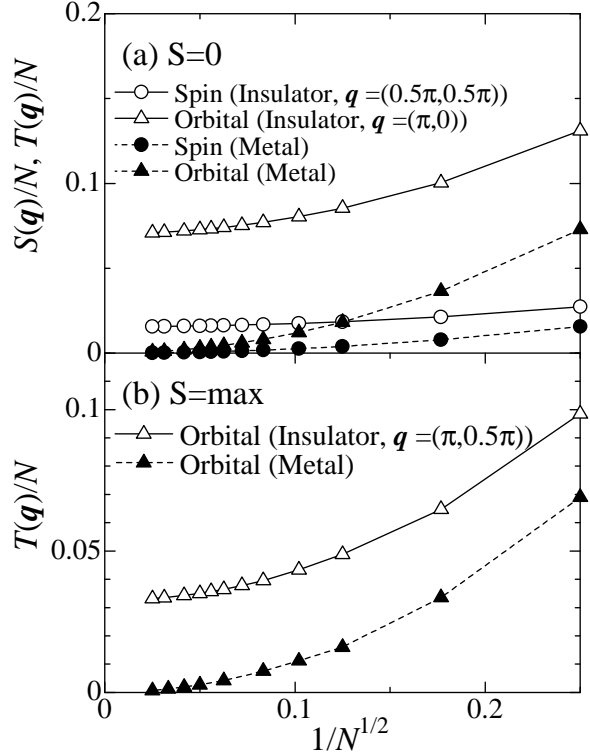}
\end{center}
\caption{
Largest spin-spin and orbital-orbital correlation functions (equal-time structure factors) by HFA as a function of $1/\sqrt{N}$ for the various wave vectors for the two-dimensional square lattices; (a) Fig. \ref{order} (a) in total $S$=0 sector at $\lambda=0.40$ (metal) and $\lambda=1.00$ (insulator). (b) Fig. \ref{order} (c) in total $S$=max sector at $\lambda=0.72$ (metal) and $\lambda=1.00$ (insulator). In each panel, circles and triangles denote the spin-spin and orbital-orbital structure factors, respectively.
}
\label{correlation_functions}
\end{figure}
Figure \ref{correlation_functions} shows peak values of the spin and orbital structure factors by HFA
when we assume $S$=0 and $S$=max states, where wavevectors ${\vec q}$ dependences of each structure factor in both metallic phases are
quite weak. With increasing system size, each correlation function obtained by the HFA monotonically decreases. All the scaled structure factors $S({\vec q})/N$ or $T({\vec q})/N$
of metallic phases in each spin sector become zero when $N \rightarrow \infty$, which indicates metallic phases become paramagnetic ($S$=0)
and paraorbital states ($S$=0 and $S$=max).
However, the magnetic and orbital orders of the insulating states appear to survive when $N \rightarrow \infty$ and the spin and orbital
long-range order may exist in ${\rm Sr_{2}VO_{4}}$ within HFA of states illustrated in Fig. \ref{order} (a) and (c).

\begin{figure}[tb]
\begin{center}
\includegraphics[width=8cm]{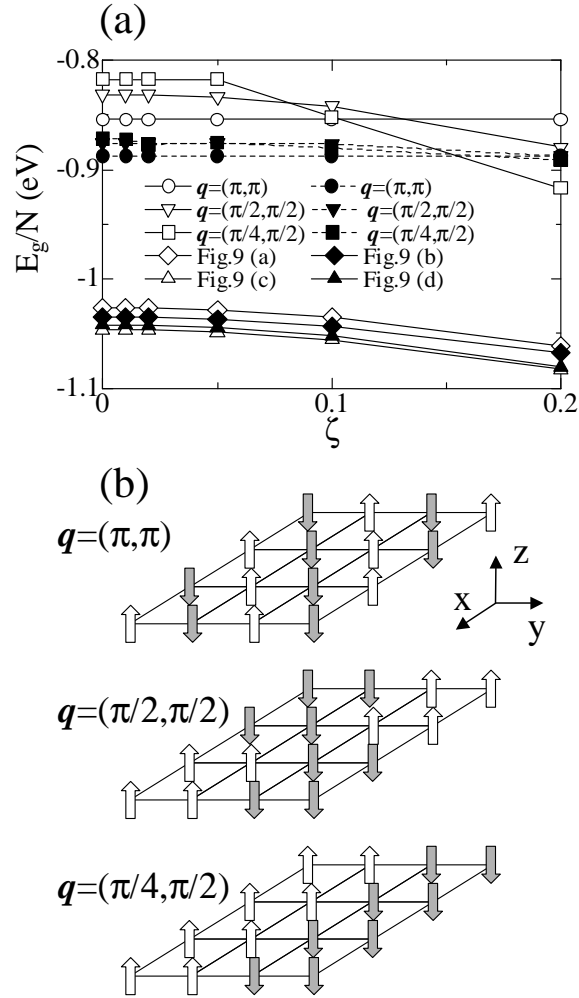}
\end{center}
\caption{(a) Ground state energies per site as a function of the spin-orbit interaction parameter $\zeta$ obtained by the HFA at $\lambda=1.00$.
The open (closed) circle, inverse triangle, and square represent structures with the spin periodicity 1$\times$1, 2$\times$2, and 4$\times$2 illustrated in (b) with $z$ $(x)$ component of spin-orbit coupling. The open and closed diamond (triangle) denote (a) and (b) ((c) and (d)) of Fig.\ref{order} with $z$ component of spin-orbit coupling, respectively.
(b) Structures of total angular momenta, which correspond to up and down arrows, at each wavevector ${\vec q}$ when spin-orbit interactions are taken into account.
}
\label{spin_orbit}
\end{figure}

In order to investigate effects of the spin-orbit coupling, we add the spin-orbit interaction term $H_{\rm SO}=\zeta\sum_{i}{\mib l}_{i}\cdot {\mib s}_{i}$ to the Hamiltonian eq.(\ref{eqn_effham}), where $\zeta$ represents the strength of the spin-orbit coupling.
Note that we first consider $z$ component of the spin-orbit coupling in this paper, for simplicity and an order parameter is defined as $\Delta^{\rm so}_{\sigma}=-{\rm i}\langle c^{\dag}_{B\sigma}c_{C\sigma}-c^{\dag}_{C\sigma}c_{B\sigma} \rangle$, where $B$ and $C$ represent $yz$ and $zx$ orbitals, respectively.
Furthermore, by rotating each axis ($x \rightarrow y$, $y \rightarrow z$, and $z \rightarrow x$), we investigate the effect of $x$ component of the spin-orbit coupling.
Within HFA, the ground state energy is shown as functions of $\zeta$ in Fig.\ref{spin_orbit}.
Note that decoupling processes of the Hamiltonian for various ordered states are different from that of the Fig. \ref{order} (a),
so that each energies at $\zeta=0.0$ do not correspond to that of Fig. \ref{order} (a).

\begin{figure}[tb]
\begin{center}
\includegraphics[width=8cm]{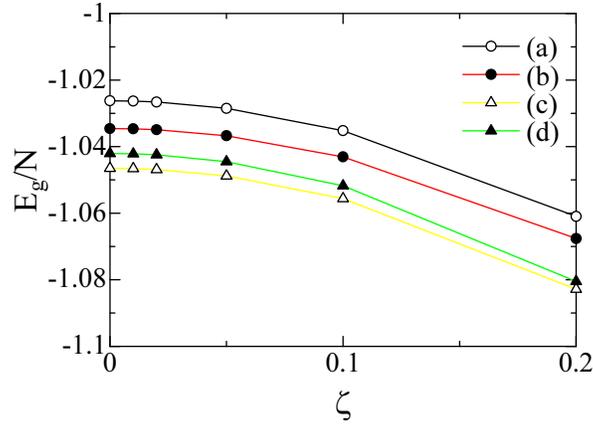}
\end{center}
\caption{(color online) Ground state energies per site as a function of the spin-orbit interaction parameter $\zeta$ obtained by the HFA at $\lambda=1.00$.
The notations from (a) to (d) have one-to-one correspondence with those in Fig.\ref{order}.}
\label{Spin-OrbitFig14}
\end{figure}

Although all the energies decrease with increasing the spin-orbit coupling, those are insensitive at $\zeta$ $<$ 0.1 eV. The realistic values of spin-orbit interactions of typical transition metals are usually 0.01$\sim$ 0.02 eV.
Even if the spin-orbit coupling is increased, the intersection of the energies of (a)-(d) in Fig.\ref{order} does not appear at $\zeta$ $<$ 0.2 eV, which is seen in Fig.\ref{Spin-OrbitFig14}. 
Furthermore the energy difference per site between the ordered state in Fig.\ref{order}(a) and other ordered states with order parameter $\Delta^{\rm so}$ is larger than 0.15eV, so that we conclude the spin-orbit coupling is irrelevant to the ground state property of ${\rm Sr_{2}VO_{4}}$. The PIRG result shown later does not alter this conclusion. Hereafter we neglect effects of the spin-orbital coupling.

The HFA results for the downfolded effective Hamiltonian of ${\rm Sr_{2}VO_{4}}$ predict that the ground state is complete ferromagnetic and insulating with the charge gap $\sim 0.2$ eV. These are not consistent with the experimental results, where the ground state appears to be an antiferromagnetic insulator.

\subsection{PIRG Results}
\subsubsection{Metal-Insulator Boundary and Orbital-Spin Structure}
In this subsection, we show PIRG results and discuss correlation effects with quantum fluctuations beyond the HFA.
Because of the layered structure, the coupling along the $z$ direction is very small. Therefore we first consider two-dimensional lattices.
Note that the interaction parameter $\lambda$ is changed at 0.08 intervals. Namely, $\lambda$=0.88, 0.96, 1.04, 1.12, $etc$. in PIRG calculations, and we carry out
interpolations by using obtained results at various $\lambda$.

\begin{figure}
\begin{center}
\includegraphics[width=9cm]{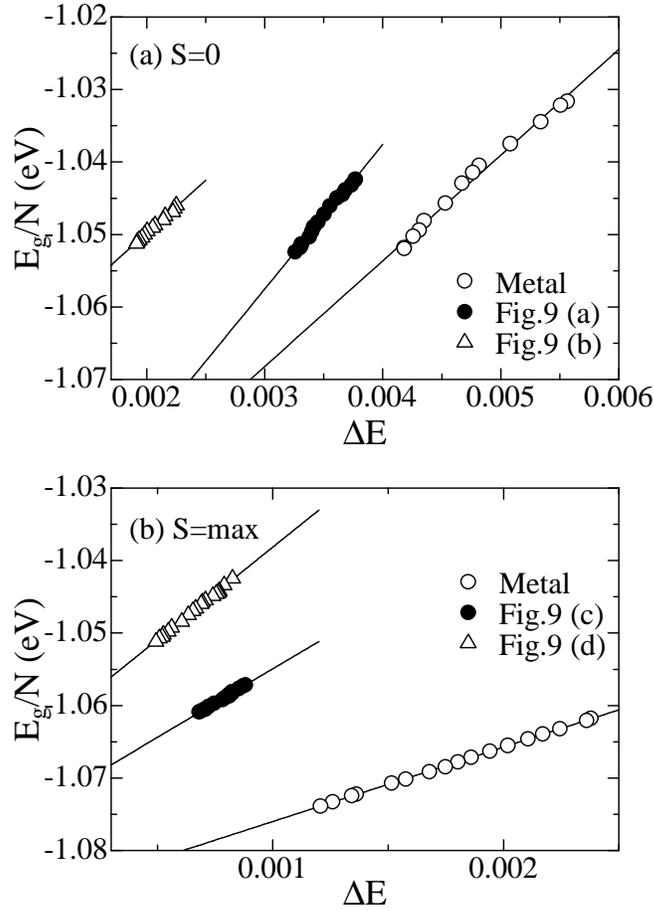}
\end{center}
\caption{
Extrapolations of the total energy per site to zero energy variance for two-dimensional 4$\times$4$\times$1 lattice at $\lambda$=1.04. $L$ is taken up to $L=192$.
In the upper panel, open circle, closed circle, and triangle represent the metal, insulator in Fig. \ref{order} (a), and insulator in Fig. \ref{order} (b), respectively at $S$=0 sector. On the other hand, in the lower panel, open circle, closed circle, and triangle represent the metal, insulator in Fig. \ref{order} (c), and insulator in Fig. \ref{order} (d), respectively at $S$=max sector.
}
\label{vari_eg}
\end{figure}

The HFA results show that the ground state belongs to the total $S$=max sector (Fig.\ref{order} (c)), whereas the ordered state in Fig.\ref{order} (b) becomes stable within the total $S$=0 sector. However, energies of other ordered states are also close to those ordered states. To estimate quantum fluctuation effects on this severe competition, we perform PIRG calculation starting from initial eigenvectors of (a)-(d) in Fig.\ref{order} and metallic states obtained by the HFA. Even after the PIRG procedure, all of these states remain metastable or stable.
The metallic states remain paramagnetic and paraorbital and ordered metallic states do not appear even after PIRG calculations.
In the large $L$ region, linear fittings against the energy variance in the $S$=0 and $S$=max sectors for various states are shown in Fig. \ref{vari_eg} at $\lambda=1.04$, for example.

Compared with results of the $S$=max sector, the linear fitting in the $S$=0 sector becomes satisfied only at large $L$ since required number of basis to obtain accurate eigenstates of the Hamiltonian generally becomes large, which means that HFA is insufficient to describe the true ground state in $S$=0 sector.
Furthermore, this behavior is especially remarkable in the metallic state.
It is because metallic phase tends to become unstable at large $\lambda$ if $L$ is increased and collapses to insulating ordered states. Even at $\lambda \sim 1.0$, though the metallic states remains metastable, the linear fitting requires larger $L$ than that of the insulating phase.

After this linear fittings, extrapolated values are obtained. Table \ref{coefficients} shows coefficients of the linear extrapolations for various states at $\lambda=1.04$. The energy holds the relation $E=s\Delta E+E_{g}$ where $s$ represents the slope of the linear extrapolation and $E_{g}$ is the ground state energy. Here note that the number of Hilbert subspace $L$ is mainly taken up to $L=192$.
\begin{table}
\caption{Coefficients of the linear extrapolation for various states at $\lambda=1.04$ ($E=s\Delta E+E_{g}$).
}
\begin{tabular}{ccccc} \hline
state($S$=0) & $s$ & $E_{g}$ & error bar \\ \hline
Metal & 13.04 &  -1.105 & $\pm$ 0.0025 \\
(a) of Fig. \ref{order} & 19.95 & -1.117 & $\pm$ 0.0013\\
(b) of Fig. \ref{order} & 14.58 & -1.079 & $\pm$ 0.0008\\ \hline \hline
state($S$=max) & $s$ & $E_{g}$ & error bar \\ \hline
Metal & 10.28 & -1.086 & $\pm$ 0.0001 \\
(c) of Fig. \ref{order} & 18.64 & -1.074 & $\pm$ 0.0003 \\
(d) of Fig. \ref{order} & 25.58 & -1.064 & $\pm$ 0.0002 \\ \hline\\
\end{tabular}
\label{coefficients}
\end{table}

Let us discuss the total energy of metallic and insulating states for each total spin sector.
First we consider the $S$=0 insulating states where HFA result predicts that the lowest state is Fig. \ref{order} (b). However, by including quantum fluctuations by PIRG, the insulating state Fig. \ref{order} (a) becomes stable in comparison with Fig. \ref{order} (b) states at $\lambda \sim 1.0$.
Since the spin configuration of Fig. \ref{order} (b) has a stripe structure with the width 2, the ferromagnetic pattern becomes dominant in the $x$ direction within HFA, so that the required number of bases to obtain the true ground state of Fig. \ref{order} (b) is smaller than that of Fig. \ref{order} (a). In fact, the energy variance of Fig. \ref{order} (b) shown in Fig. \ref{vari_eg} is small in comparison with that of Fig. \ref{order} (a). Therefore, though the energies at the same $L$ are comparable between (a) and (b), Fig. \ref{order} (a) has lower energy in insulating state at $S$=0 sector after the extrapolation to zero energy variance, where the total energy $E_{g}=-1.117 \pm 0.0013$ is obtained at $\lambda$=1.04.
Furthermore, compared with the total energy of the metallic state, where $E_{g}=-1.105 \pm 0.0025$, we can conclude that the insulating state Fig. \ref{order} (a) becomes most stable in $S$=0 sector.

On the other hand, in $S$=max sector, HFA result predicts that the insulating state Fig. \ref{order} (c) has lowest energy.
In the insulating states, the total energy of Fig. \ref{order} (c) obtained by PIRG, which is $E_{g}=-1.074$, becomes lower than that of Fig. \ref{order} (d). However, compared with the metallic state, Fig. \ref{order} (c) becomes a higher energy state. Therefore the metallic state becomes most stable in $S$=max sector, where $E_g=-1.086$ at $\lambda$=1.04.

At $\lambda \sim 1.0$, in the $S$=0 sector, the ground state turns out to be insulating and ordered as (a) in Fig. \ref{order} after the extrapolation to zero energy variance. On the other hand, although Fig. \ref{order} (c) becomes most stable in the insulating states, the ground state becomes metallic in the $S$=max sector. Therefore, hereafter we only consider the configurations Fig. \ref{order} (a) at total $S$=0 and Fig. \ref{order} (c) at total $S$=max in the insulating states.

\begin{figure}[tb]
\begin{center}
\includegraphics[width=8cm]{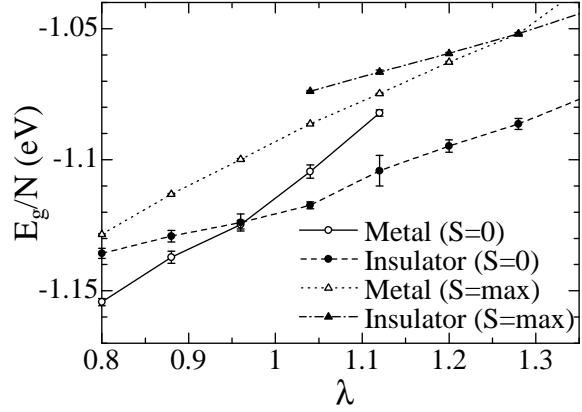}
\end{center}
\caption{Total energies as a function of interaction $\lambda$ by PIRG calculations: total $S$=0 and $S$=max states for $4\times$4$\times$1 lattice, where insulating phases in each spin sector correspond to Fig. \ref{order} (a) and (c), respectively.
}
\label{energy}
\end{figure}

Next, let us consider the metal-insulator transitions in each total spin sector. The total energy per site as a function of the interaction $\lambda$ is shown in Fig. \ref{energy} for $4\times$4$\times$1 lattice.
Although the energy of total $S$=max obtained by HFA is lower than that of $S$=0, the PIRG predicts opposite around $\lambda \sim 1.0$.
The energies of metallic and insulating states intersect at $\lambda \sim 1.0$ (for the $S$=0 state) and at $\lambda \sim 1.3$
(for the complete ferromagnetic state), where the metal-insulator transitions occur. In particular, total $S$=0 state is close to
the metal-insulator transition, which indicates the realistic parameter for ${\rm Sr_{2}VO_{4}}$ is on the verge of the Mott transition.

\begin{figure}[tb]
\begin{center}
\includegraphics[width=8cm]{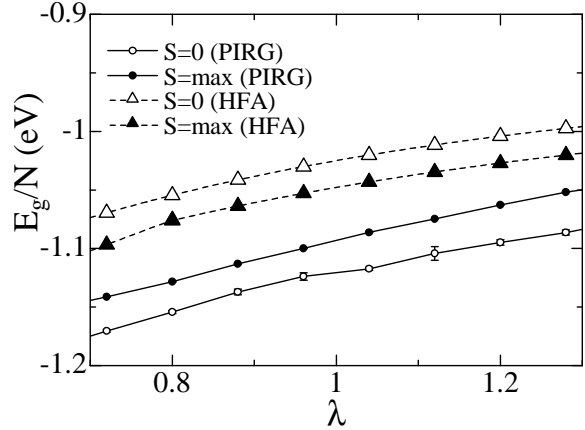}
\end{center}
\caption{Lowest energies per unit cell as a function of interaction $\lambda$ by HFA and PIRG for total $S$=0 and complete ferromagnetic states.
}
\label{eg_hf_pirg}
\end{figure}

In order to consider the energy gain in comparison with HFA, we show the lowest total energies obtained by PIRG for each spin sector as a function of interaction $\lambda$ in Fig. \ref{eg_hf_pirg} ((a) at total $S$=0 and (c) at complete ferromagnetic state in Fig. \ref{order}) .
The difference between the ground state energy of total $S$=0 and lowest energy  of total $S$=max states is 0.03eV around $\lambda \sim 1.0$ within PIRG results.
We therefore conclude that the true ground state becomes total $S$=0 state in contrast to the result of HFA prediction.
The available experimental results are consistent with our obtained result, where ${\rm Sr_{2}VO_{4}}$ shows antiferromagnetic Mott insulator with small gap.

Since the candidates of configuration for spin and orbital order at $S$=0 are in severe competition within 0.01 eV at $\lambda \sim 1.0$, where the total energy of other ordered state is close to that of (a) in Fig. \ref{order} (not shown), we can not exclude the possibility of other ordered states.
However, all the candidates have large unit cells with nontrivial lattice structures.

The obtained results are qualitatively different from the HFA results. The origin of this discrepancy is interpreted as follows; It is well known that the ordered state tends to be overestimated in the HFA since quantum fluctuations are not included. By taking account of quantum fluctuations by PIRG,
the critical values of interactions for ordered states become large.
In particular, the energy gain by quantum fluctuations is conspicuous in $S$=0 state than the ferromagnetic state. Therefore the required number of bases becomes larger in comparison with that of the complete ferromagnetic state to obtain an accurate estimate of the ground state, so that the energy gains of $S$=0 sector tends to become large with increasing the Hilbert subspace $L$.
Therefore the single Slater determinant approximations, such as HFA and LDA+U approach, can not describe the antiferromagnetic states.

On the other hand, energy gains at total $S$=max by the quantum fluctuations are rather small because ferromagnetic configurations
may be relatively well given by a single Slater determinant.  In general, the ferromagnetic state is represented by fewer number of basis functions and quantum fluctuations are small.

\begin{figure}[tb]
\begin{center}
\includegraphics[width=8cm]{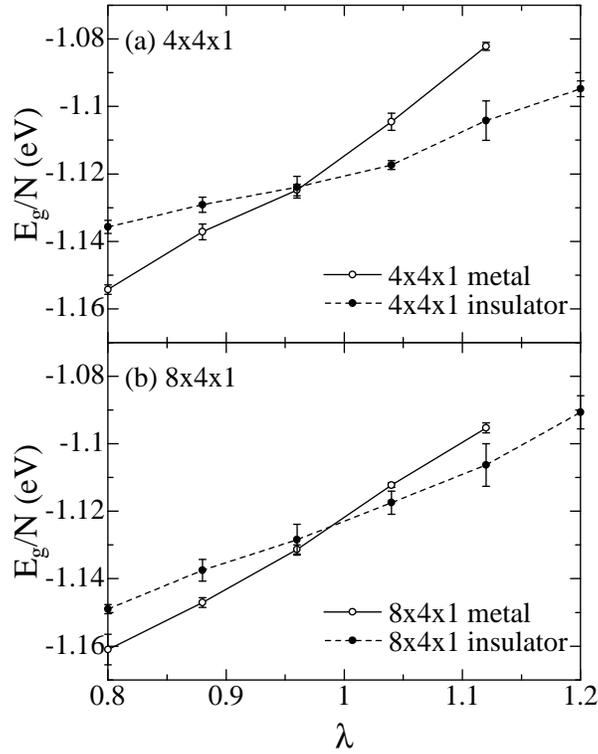}
\end{center}
\caption{Total energies as a function of interaction $\lambda$ for $4\times$4$\times$1 and $8\times$4$\times$1 lattices in total $S$=0 sector.
}
\label{mit2}
\end{figure}

We conclude that ${\rm Sr_{2}VO_{4}}$ is close to the Mott transition and may have complicated spin and orbital order with Fig. \ref{order} (a) structure for $4\times$4$\times$1 lattice.
Here we investigate the system size dependence of metallic and insulating phases illustrated in Fig. \ref{order} (a) in $S$=0 sector.
In Fig. \ref{mit2}, we show PIRG results of total $S$=0 for different lattice sizes. This result for $8\times$4$\times$1 lattice is similar to that of $4\times$4$\times$1 lattice, where insulating state for two system sizes becomes stable in comparison with metallic states and the metal-insulator transition occurs at $\lambda \sim 1.0$. Therefore we believe that size dependence is small.

Furthermore, in order to consider the three dimensional effect, we have calculated the total energy of some three-dimensional ordered patterns with the order (Fig. \ref{order} (a)) on two-dimensional plane being fixed.
However the difference of each energy is very small (within 0.001eV). Since the ratio of magnitudes of inter-layer hoppings and intra-layer hoppings is smaller than $1/10$, the inter-layer correlation is quite weak.
In addition, frustration is induced by ferromagnetic and antiferromagnetic interactions between layers. Therefore the ground state has severe degeneracy, so that we can not determine the unique spin-orbital ordering structure in three-dimensional system within the present study.
It is even conceivable that the interlayer correlation is frozen to short-range pattern in realistic experimental conditions.

Here we refer to the magnetic and orbital orders. We have calculated structure factors eq. (\ref{eqn_cf_spin}) and (\ref{eqn_cf_orbital}) within the PIRG for 4$\times$4$\times$1 and 8$\times$4$\times$1 lattices. Although correlation functions have larger error bars than the estimate of the ground state energy in PIRG and the accuracy of the estimate of the long-range order becomes worse, the amplitude of the long-range order does not have serious system size dependence and similar to the size dependence obtained in HFA.
Therefore we believe that the magnetic and orbital orders of Fig. \ref{order} (a) survives in the thermodynamic limit.

\begin{figure}[bu]
\begin{center}
\vspace{-5mm}
\includegraphics[width=8cm]{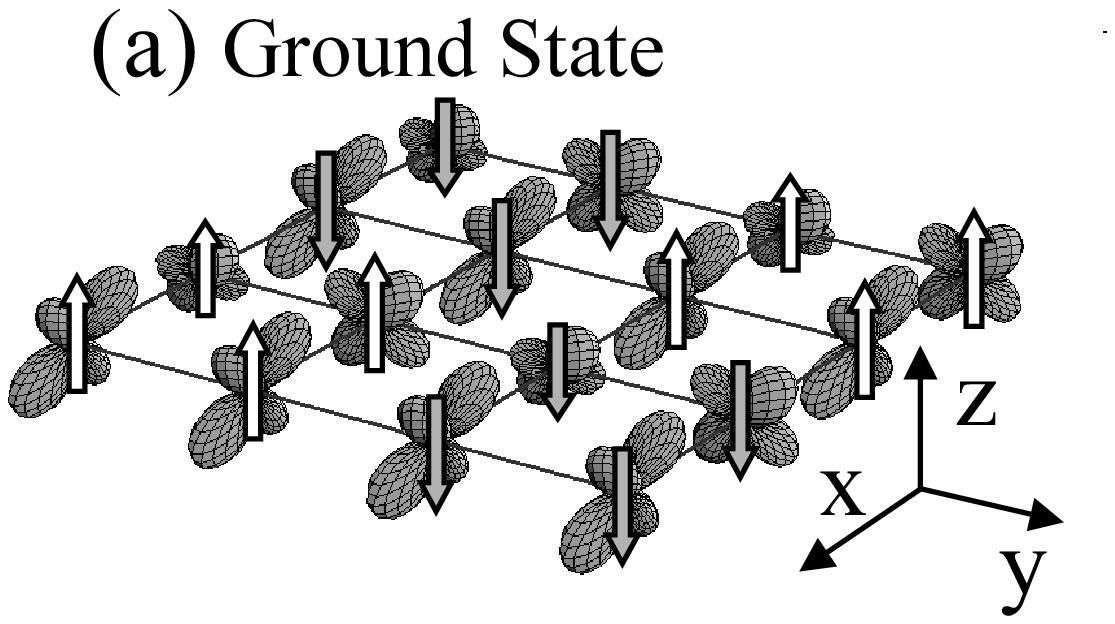}
\vspace{-5mm}
\includegraphics[width=8cm]{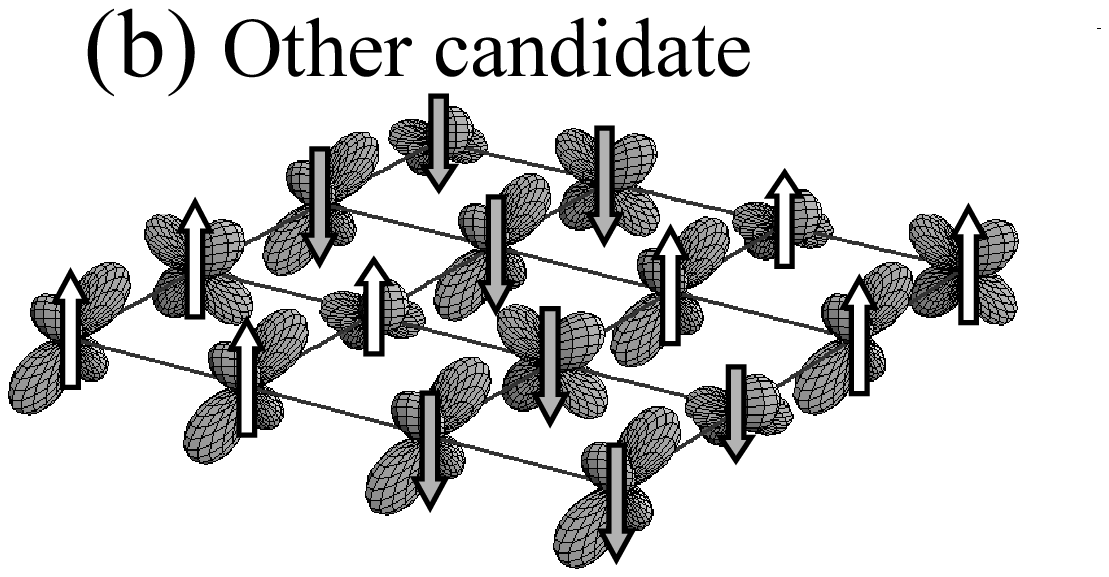}
\end{center}
\caption{Obtained spin-orbital structres when $\lambda \sim 1.0$ at $L$=160.
(a) and (b) correspond to the ground state and one of the other candidate, respectively.
}
\label{pirg_order}
\end{figure}

Figure \ref{pirg_order} shows the spin-orbital structures calculated by PIRG. (a) represents the ground state, which is obtained by using the HFA result (Fig. \ref{order} (a)) as a initial state. In comparison with HFA structure, minor orbital component rather than the dominant orbital at each site becomes larger by including quantum fluctuations in PIRG, so that obtained orbitals rotate from each axis. On the other hand, figure \ref{pirg_order} (b) is one of the other candidates. The energy difference between (a) and (b) is in the order of 0.01eV, where low energy states are in severe competition.

\begin{figure}[tb]
\begin{center}
\includegraphics[width=8cm]{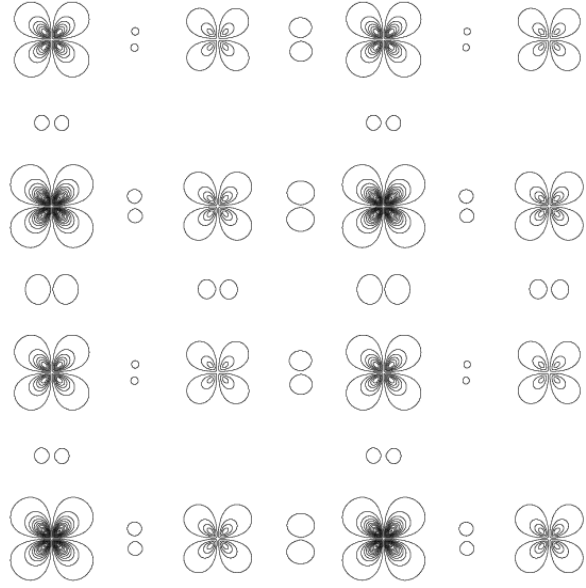}
\end{center}
\caption{Contour plot of the charge density of vanadium `` $t_{2g}$'' Wannier orbitals in the mirror $ab$ plane consisting of V and O, constructed for the 4$\times$4$\times$1 unit cell at $\lambda = 1.04$ for number of basis at $L=160$. In these contour lines, the minimal value of the charge density shown is 0.004, which increases with the interval 0.022. For certain oxygen sites the maximal intensity appears to be smaller than 0.004 and are not explicitly visible (blank spaces on the figure).
}
\label{charge_density}
\end{figure}

Since the wave function can be directly calculated in PIRG, we can straightforwardly calculate the charge density in the realistic system after transforming to the real space representation of the Wannier orbitals of each `` $t_{2g}$'' electrons.
Figure \ref{charge_density} shows the charge density of unit cell in ${\rm Sr_{2}VO_{4}}$ for each number of basis state $L$ at $S$=0 state in the VO$_2$ plane at $z=0$, where the insulating state is indeed realized at $\lambda$=1.0 (shown in Fig. \ref{mit2}).

\subsubsection{Optical Conductivity}
Recently the optical conductivity was measured by Matsuno {\it et al.}\cite{mats03,mats05}. Here we discuss the optical conductivity for the spin-orbital ordered state obtained by PIRG.

Since it is difficult to evaluate dynamical quantities directly within PIRG, to calculate the conductivity, we complementarily use a HFA eigenstate which simulates the PIRG results instead of the PIRG wavefunction itself.
In the PIRG results, order parameters obtained by HFA are quantitatively reduced because of quantum fluctuations.
Therefore, though it is not complete, we employ the solution of HFA which reproduces the peak amplitude of the order-parameter structure factor of the PIRG state, by taking reduced interaction parameters.
In order to obtain this approximate ground state, we consider the additional procedure in HFA. We first employ the eigenstate of HFA at $\lambda=1.0$, $|\Phi\rangle$, which is the metastable state with the same spin-orbital configuration as the PIRG ground state.  Namely, we employ the Hartree-Fock state shown in Fig.\ref{order}(a).  Then by introducing new interaction parameters $\lambda'_{xy}$, $\lambda'_{yz}$, and $\lambda'_{zx}$, we newly calculate matrix elements $\langle \Phi|H_{\rm eff}|\Phi\rangle$, where the interaction part of $H_{\rm eff}$ is replaced as,
\begin{eqnarray}
 & &U'_{xy,yz}\Big(\langle n_{xy} \rangle n_{yz}
                 + n_{xy} \langle n_{yz} \rangle \Big) \nonumber \\
 &+&U'_{yz,zx}\Big(\langle n_{yz} \rangle n_{zx}
                 + n_{yz} \langle n_{zx} \rangle \Big) \nonumber \\
 &+&U'_{zx,xy}\Big(\langle n_{zx} \rangle n_{xy}
                 + n_{zx} \langle n_{xy} \rangle \Big) \nonumber \\
&\rightarrow &
   \lambda'_{xy}\Big(U'_{xy,yz}\langle n_{yz} \rangle
                    +U'_{zx,xy}\langle n_{zx} \rangle \Big)n_{xy}\nonumber \\
&+&\lambda'_{yz}\Big(U'_{yz,zx}\langle n_{zx} \rangle
                    +U'_{xy,yz}\langle n_{xy} \rangle \Big)n_{yz}\nonumber \\
&+&\lambda'_{zx}\Big(U'_{zx,xy}\langle n_{xy} \rangle
                    +U'_{yz,zx}\langle n_{yz} \rangle \Big)n_{zx}.
\label{reduced_HFA}
\end{eqnarray}
Here, $U'$ represents inter-orbital interactions $K-J$ or $K$.
By diagonalizing the modified HFA Hamiltonian, new eigenstates $|\Psi_{n}\rangle$ can be obtained. By optimizing the parameters $\lambda'_{xy}, \lambda'_{yz}$ and $\lambda'_{zx}$, we seek for the Hartree-Fock ground state $|\Psi_0\rangle$ which reproduces relevant physical quantities obtained by PIRG.

As the relevant physical quantities, we employ the charge-charge correlation functions for each orbital $m$, which are defined as
\begin{eqnarray}
C_{m,m}({\mathbf q})=
\frac{\pi}{N}\sum_{i,j}\langle n_{im} n_{jm} \rangle e^{{\rm i} {\mathbf q}\cdot ({\mathbf R}_{i}-{\mathbf R}_{j})},
\end{eqnarray}
where $n_{im}=n_{im\uparrow}+n_{im\downarrow}$. The largest values appear at ${\mathbf q}=(\pi,0)$ in all correlation functions in HFA and PIRG (shown in Table \ref{cccf}), whose results show that quantum fluctuations suppress each amplitude of correlation functions.
\begin{table}
\caption{The largest values of charge-charge correlation functions obtained by HFA and PIRG at ${\mathbf q}=(\pi,0)$ at $\lambda=1.0$ for 4$\times$4$\times$1 lattices.
}
\begin{tabular}{ccc} \hline
           & HFA  &  PIRG  \\ \hline
$C_{xy,xy}({\mathbf q})$ & 0.45 &  0.22  \\
$C_{yz,yz}({\mathbf q})$ & 0.98 &  0.97  \\
$C_{zx,zx}({\mathbf q})$ & 3.1 &  2.3  \\ \hline
\end{tabular}
\label{cccf}
\end{table}

When $\lambda'_{xy}=0.59$, $\lambda'_{yz}=0.90$, and $\lambda'_{zx}=0.62$ in eq. (\ref{reduced_HFA}), the peak values obtained by the modified HFA nearly coincide with the PIRG results. Therefore we regard this eigenstate obtained by the modified HFA as an approximate ground state.

Then we calculate optical conductivity using this Hartree-Fock results.  The optical conductivity is defined as,
\begin{eqnarray}
\sigma_{\alpha\alpha}(\omega)=\frac{\pi}{N\omega}\sum_{M\ne 0}
\frac{|\langle \Psi_{M}|j_{\alpha}|\Psi_{0}\rangle |^{2}}{\omega -E_{M}+E_{0}},
\end{eqnarray}
where $\Psi_{0}$($\Psi_{M}$) is the ground (excited) state with the total energy $E_{0}$ ($E_{M}$) obtained by the above mentioned procedure. The current operator $j_{\alpha}$ along $\alpha$ direction is written as
$j_{\alpha}=-{\rm i}\sum_{i,m,m'\sigma}t^{mm'}_{i,i'_{\alpha}}
(c^{\dag}_{im\sigma}c_{i'_{\alpha}m'\sigma}
-c^{\dag}_{i'_{\alpha}m'\sigma}c_{im\sigma})$
where $i'_{x}=(i_{x}+1,i_{y})$ and $i'_{y}=(i_{x},i_{y}+1)$.

\begin{figure}[h!]
\begin{center}
\includegraphics[width=8cm]{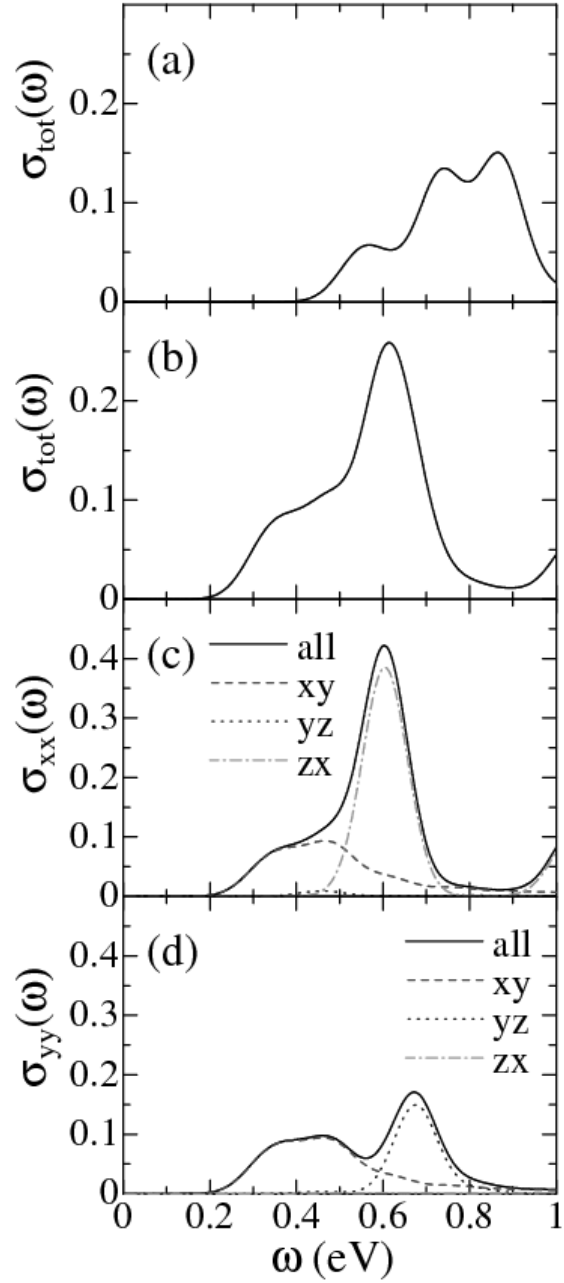}
\end{center}
\caption{
Optical conductivity obtained by (a) HFA and (b-d) modified HFA at $\lambda=1.0$, $\lambda'_{xy}=0.59$, $\lambda'_{yz}=0.90$, and $\lambda'_{zx}=0.62$ to simulate the PIRG result. 
Panels (a) and (b) are total conductivities, $\sigma_{\rm tot}(\omega)=(\sigma_{xx}(\omega)+\sigma_{yy}(\omega))/2$ obtained by assuming the twin structure of orbital-spin order obtained by the configuration of Fig.\ref{pirg_order}(a) mixed with the 90 degree rotated structure. Panels (c) and (d) show $\sigma_{xx}(\omega)$ ($\sigma_{yy}(\omega)$), where solid line is full conductivity and dashed, dotted, and dashed-dotted lines represent contributions of $xy$, $yz$, and $zx$ orbitals.}
\label{opt}
\end{figure}

Figure \ref{opt} shows optical conductivities at low energy regime ($\omega < 1$ eV, (a) HFA and (b-d) modified HFA), which consists of the contribution of excitations in the inter-orbital interactions $K-J$ terms.
In the higher energy region in (a), although there are other peaks caused by the scattering of the inter-orbital interactions ($K$ terms, $\omega \sim 1.5$eV) and intra-orbital interactions ($U$ terms, $\omega \sim 3.0$eV), the amplitude is relatively smaller than that of $K-J$ terms. Therefore these peaks are not essential in our discussion. The $K-J$ peak basically represents transitions of an electron to an orbital $m$ on the neighboring sites where another electron with the parallel spin already exists at the different orbital from $m$. This process makes the energy cost of inter-orbital onsite interaction $K$ reduced by the Hund's rule coupling $J$. 
The total conductivity (b) is the obtained by optimizing the parameters $\lambda'_{xy}$, $\lambda'_{yz}$, and $\lambda'_{zx}$, where peak positions shift to the low energy region in comparison with the HFA result (a). 
This result may approximately represent the realistic conductivity of ${\rm Sr_{2}VO_{4}}$. 

In order to clarify contributions from each orbital in the $K-J$ peak of the optical conductivity, the transfers except for one orbital are switched off in current operator (Fig. \ref{opt} (c) and (d) ).
Note that the amplitude of sharp peak in $\sigma_{xx}(\omega)$ ($\omega \sim 0.65$ eV) is  approximately twice as large as that in $\sigma_{yy}(\omega)$ because the spin-orbital structure on Fig. \ref{pirg_order} (a) has the anisotropy about the direction. Since the orbital ferromagnetic structure to $y$ direction, which consists of $zx$ orbitals, is partially realized, the current along $y$ direction is reduced. 
While spectra of the $xy$ orbital are broad and have a diminished gap, those of the $zx$ orbital in $\sigma_{xx} (\omega)$ and $yz$ orbital in $\sigma_{yy} (\omega)$ have sharp peak structures because the $xy$ band is broader than $yz$ and $zx$ orbitals. We note that the gap of the band mainly arising from the $xy$ orbital is significantly reduced as compared with that of the HFA at $\lambda=1.0$. This selective reduction of the $xy$ gap makes a shoulder at $\omega \sim 0.40$eV and a peak at $\omega \sim 0.65$eV in the optical conductivity at low energy region, whose overall structure is consistent with the experimental result\cite{mats05}.  We note that it is difficult to explain the shoulder structure without the present clarification.

There exist some disagreement of peak positions between the present result and the experimental result.  In the experiment, the peak position is around 1 eV, while our result shows peak at smaller energy.  Such discrepancy may be ascribed to the oversimplification of the PIRG result by the single Slater determinant in the imitated HFA.
In general, even if the long range order disappears, the insulating gap does not always vanish. 1D Hubbard model is a typical example. However, within the HFA, when the long range order disappears, it always become a metal. When we reproduce the correct order parameters by optimizing the parameters $\lambda'_{xy}$, $\lambda'_{yz}$, and $\lambda'_{zx}$ within the Hartree-Fock approximation, it ignores the gap formation through the genuine Mott origin without the long-range order. Then it may shift the conductivity gap to too low energy region, 
which may be the main reason of disagreement of peak positions. 

When the transfer is increased relative to the interaction from the experimental condition, (namely $\lambda$ is reduced), for example, by pressure, the present result opens an interesting possibility of orbital selective metallization, where electrons with $xy$ component becomes metallic while the other component keeps the gap structure.

Orbital selective Mott transitions (OSMT) have extensively been studied in systems with orbital degeneracy for $d^2$ system.  Theoretically, dynamical mean-field theory appears to support that OSMT takes place when the Hund exchange term is included and the orbital hybridization is absent~\cite{Kawakami}.  Experimentally, (Ca,Sr)$_2$RuO$_4$ has been proposed to be a good candidate for the occurrence of the OSMT~\cite{Rice,Maeno}.  However, (Ca,Sr)$_2$RuO$_4$ has a substantial tilting of the octahedra (so called GdFeO$_3$-type distortions) in the perovskite structure because of the large ionic radius of Ru atom. This tilting yields substantial hybridization between orbitals, which could mask the real OSMT. In fact, theoretically it was suggested that the OSMT would be sensitively destroyed by the orbital hybridization.

In Sr$_2$VO$_4$, the tilting is absent and therefore the hybridization between orbital can be neglected.
In fact, recent optical conductivity data~\cite{mats05} are successfully explained if only one orbital is on the verge of the Mott transition while other orbitals keep large gap, which is indeed seen in our DFT-PIRG result.
A crucial difference from (Ca,Sr)$_2$RuO$_4$ is, however, that Sr$_2$VO$_4$ is $d^1$ system, where at the metal insulator transition of $xy$ orbital, other orbitals are basically empty. In other words, if the $xy$ component of electrons is metallized, other component is not able to keep the commensurability condition strictly required for the Mott insulator.  Therefore we do not expect the original sense of OSMT anyhow.  However, in the analogy of OSMT, it would be conceivable to keep an insulating gap for the excitation of electrons in the $yz$ and $zx$ orbitals, if we expect a continuous evolution of the optical spectra. (Note that the $d^1$ electron in the ground state has substantial amount of $yz$ and $zx$ component in our PIRG result.)  Our DFT-PIRG result, however, shows that such intermediate phase may be preempted by a single first-order Mott transition with concurrent vanishing of all the gaps of the orbitals in this realistic case.  Namely, right after the transition to the metallic state, the PIRG result predicts that the orbital-spin order collapses and the whole gap structure vanishes immediately.

Although DMFT supports the existence of OSMT for $d^2$ system, DMFT has a limitation of ignoring the spatial correlation.  The present result implies that, even in the $d^2$ systems, spatial fluctuations and collapse of orbital-spin order structure may have a serious effect on the stability of the intermediate phase at OSMT. It would be an intriguing open issue whether OSMT may theoretically survive in the realistic situation, where the insulating gap is substantially influenced by the spatial structure of the orbital-spin order. A possible situation is that the collapse of one component orbital order at the metallization may destroy other order cooperatively, which results in the simultaneous collapse of the whole gap.

\begin{figure}[h!]
\begin{center}
\includegraphics[width=8cm]{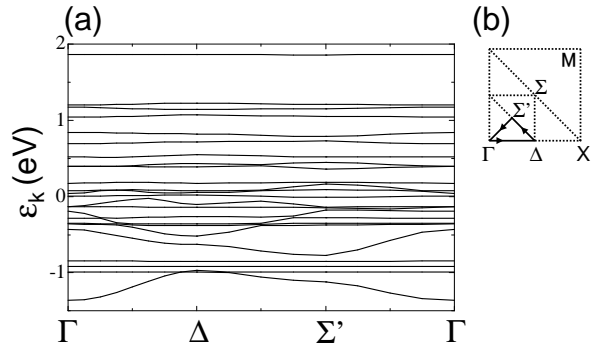}
\end{center}
\caption{One-particle energy dispersions at $\lambda=1.0$, $\lambda'_{xy}=0.59$, $\lambda'_{yz}=0.90$, and $\lambda'_{zx}=0.62$ in the reduced Brillouin zone of $4\times 2$ structure. Fermi level is at $E_{\rm F}\sim -0.8$eV.
}
\label{ek}
\end{figure}

For reference, we consider one-particle energy dispersions in our modified HFA to simulate PIRG result. The one-particle wave function in the momentum space is defined as,
\begin{eqnarray}
|{\mathbf k},n \rangle =\frac{1}{\sqrt{N}} \sum_{i}
|i,n \rangle e^{-{\rm i} {\mathbf k}\cdot (R_{i}+r_{n})},
\end{eqnarray}
where $n$ is the sublattice index (4$\times$2 (structure) $\times$ 3 (orbitals) $=24$). $R_{i}$ denotes the sublattice site and $r_{n}$ is the position in the sublattice. By using the wave function in the momentum space, the matrix element of the modified HF Hamiltonian with ${\mathbf k}$ space is given by,
\begin{eqnarray}
\epsilon^{n,n'}_{\mathbf k}&=&
\langle {\mathbf k},n|H|{\mathbf k},n' \rangle \nonumber \\
&=&\frac{1}{N} \sum_{i,j} \langle i,n|H|j,n' \rangle
e^{{\rm i} {\mathbf k}\cdot (R_{i}+r_{n}-R_{j}-r_{n'})}.
\end{eqnarray}
Diagonalizing this $24\times 24$ matrix for each ${\mathbf k}$, one-particle energy dispersions can be determined.
The obtained dispersions are shown in Fig.\ref{ek}.
The lowest 4 bands are fully occupied and other bands are empty, so that the small energy gap appears at $\sim$ $-0.8$ eV, which is reflected in optical conductivity.
Note that 3 bands in the fully occupied bands become quite flat because of the gap opening caused by the strong Coulomb interactions.

In summarizing the PIRG results, antiferromagnetic fluctuations are larger than ferromagnetic fluctuations and these fluctuations ignored in the HFA are taken into account by PIRG, which leads to a larger energy lowering of the $S$=0 state by PIRG.  The present PIRG result suggests that this energy lowering by the quantum fluctuations is crucial in determining the true ground state.  We conclude the ground state becomes the insulating state with small gap at $S$=0 at $\lambda \sim 1$ as seen from Fig.\ref{energy} with a nontrivial spin-orbital structure illustrated in Fig.\ref{pirg_order}.

\section{Summary and Discussion}
In summary, we have investigated the ground state properties of ${\rm Sr_{2}VO_{4}}$ by employing new algorithm, LDA combined with PIRG. While the high energy band structure is determined by the LDA, the electrons close to the Fermi level are sensitive to quantum fluctuation effects and are studied by more accurate computational method. To bridge these two energy scales, we have derived the low-energy effective Hamiltonian from the high-energy LDA band structure by a downfolding scheme using combinations of the constrained LDA and the GW approximation.
Since the isolated bands close to the Fermi level in ${\rm Sr_{2}VO_{4}}$ in LDA mainly consist of V $3d$ $t_{2g}$ orbitals, we extract these low-energy bands from the LDA bands in order to construct the effective Hamiltonian. Obtained effective Hamiltonian is treated by PIRG.

Applying the HFA and PIRG schemes to the obtained effective Hamiltonian for ${\rm Sr_{2}VO_{4}}$, we have calculated the total energy and physical quantities for metallic and various insulating ordered states for $S$=0 and complete ferromagnetic states.
The HFA result suggests that ${\rm Sr_{2}VO_{4}}$ becomes a complete ferromagnetic insulator with a large charge gap $\sim$ 0.2 eV.
LDA+U prediction has similar results to HFA.
On the other hand, PIRG result shows that $S$=0 state becomes the ground state, where the large unit cell structure with complicated spin and orbital pattern is realized. This complicated structure is resulted from frustration effects due to effective long-range exchanges. Furthermore ${\rm Sr_{2}VO_{4}}$ is close to the Mott transition, which implies that the charge gap is quite small.
These obtained results are consistent with available experiments in contrast to the expectations of the LDA and HFA.

While ferromagnetic states are well described by single Slater determinant approximations, such as LDA, HFA, {\it et al.}, antiferromagnetic states with large quantum fluctuations are not. By considering strong quantum fluctuations by PIRG scheme beyond a single Slater determinant,  the accurate ground state can be described.

This material shows severe competitions. In addition to being close to the Mott transition, the energy differences between various complicated ordered states at $S$=0 sector are rather small.
Although we can not exclude other configurations of magnetic and orbital order,  the ordered state certainly has large unit cell structure with a nontrivial periodicity concerning spin and orbital pattern.
It would be desired to examine the spin-orbital order proposed in this paper in careful experimental studies.

Since the Mott gap is very small, it would be interesting to apply pressure, because the gap may vanish and be metallized under reasonable and accessible range.  Although the Mott transition through the bandwidth control route can be observed in many organic conductors, there exist not many examples in inorganic systems including transition metal compounds.  If a large single crystal may be grown, it will greatly serve in clarifying the nature of the Mott transition since  sample size of the organic conductors is in most cases too small to allow various experimental studies such as the neutron scattering. A large sample may serve in clarifying the nature of Mott transition in this route, where unusual criticality is proposed~\cite{Imada2005}.  Many of the organic conductors have superconducting phases near the Mott transition while the superconductivity is absent in metals close to the bandwith control Mott transition for the transition metal compounds such as V$_2$O$_3$ under pressure and LaTiO$_3$ with oxygen deficiency. Studies of metallic phases expected for Sr$_2$VO$_4$ under pressure will contribute in clarifying the mechanism of superconductivity observed near the Mott transition including the cuprate superconductors. Another intriguing point is large orbital degeneracy and fluctuations possibly expected in the metallic phase of this material, which may have different aspect from other cases.
In fact, our result for the optical conductivity and the comparison with the experimental result show that, in the insulating side, the insulating gap is selectively reduced for electrons on the $xy$ orbital while the gap of electrons on the $zx$ and $yz$ orbitals remain. This is reminiscent of the orbital selective Mott transition recently proposed for (Sr,Ca)$_2$RuO$_4$.

\section*{Acknowledgment}
The authors would like to thank Igor Solovyev for providing us with the data for the downfolded Hamiltonian and valuable discussions. They are indebted to J. Akimitsu, Y. Tokura and J. Matsuno for illuminating discussions. They also acknowledge valuable discussions with T. Mizusaki. YI is indebted to T. Saso, S. Watanabe, and Y. Otsuka for useful discussions. A part of the numerical computations were done at the supercomputer center at Institute for Solid State Physics, University of Tokyo. A part of this work is supported by a Grant-in-aid from the Ministry of Education, Culture, Sports, Science and Technology under the grant number 16340100 and 17064004.


\begin{thebibliography}{99} 
%
\bibitem{imad98}
M. Imada, A. Fujimori and Y. Tokura:
Rev. Mod. Phys. {\bf 70} (1998) 1039.

\bibitem{cyro90}
M. Cyrot, B. Lambertandron, J. L.Soubeyroux, M. J. Rey, P. Dehauht, F. Cyrotlackmann, G. fourcaudot, J. Beille and J. L. Tholence:
J. Solid State Chem. {\bf 85} (1990) 321.
%
\bibitem{rey90}
M. J. Rey, P. Dehaudt, J. C. Joubert, B. Lambertandron, M. Cyrot and F. Cyrotlackmann:
J. Solid State Chem. {\bf 86} (1990) 101.
%
\bibitem{mats90}
S. Matsuda, S. Takeuchi, A. Soeta, T. Doi, K. Aihara and T. Kamo:
Jpn. J. Appl. Phys. {\bf 29} (1990) 1781.
%
\bibitem{noza91}
A. Nozaki, H. Yoshikawa, T. Wada, H.Yamauchi and S. Tanaka:
Phys. Rev. B {\bf 43} (1991) 181.
%
\bibitem{gian95}
V. Giannakopoulou, P. Odier, J. M. Bassat and J. P. Loup:
Solid State Commun. {\bf 93} (1995) 579.
%
\bibitem{mats05}
J. Matsuno, Y. Okimoto, M. Kawasaki and Y. Tokura:
Phys. Rev. Lett. {\bf 95} (2005) 176404.
%
\bibitem{mats03}
J. Matsuno, Y. Okimoto, M. Kawasaki and Y. Tokura:
Appl. Phys. Lett. {\bf 82} (2003) 194.
%
\bibitem{imai05}
Y. Imai, I.V. Solovyev and M. Imada:
Phys. Rev. Lett. {\bf 95} (2005) 176405.
%
\bibitem{imad00}
M. Imada and T. Kashima:
J. Phys. Soc. Jpn. {\bf 69} (2000) 2723.
%
\bibitem{kash01}
T. Kashima and M. Imada:
J. Phys. Soc. Jpn. {\bf 70} (2001) 2287.
%
\bibitem{mori02}
H. Morita, S. Watanabe and M. Imada:
J. Phys. Soc. Jpn. {\bf 71} (2002) 2109.
%
\bibitem{wata03}
S. Watanabe:
J. Phys. Soc. Jpn. {\bf 72} (2003) 2042.
%
\bibitem{mizu04}
T. Mizusaki and M. Imada:
Phys. Rev. B {\bf 69} (2004) 125110.
%
\bibitem{solo04}
I.V. Solovyev:
Phys. Rev. B {\bf 69} (2004) 134403. 
%
\bibitem{solo05a}
I.V. Solovyev and M. Imada: 
Phys. Rev. B {\bf 71} (2005) 045103. 
%
\bibitem{solo05b}
I.V. Solovyev:
cond-mat/0506632.
%
\bibitem{LMTO}
O.~K.~Andersen: Phys.~Rev.~B~{\bf 12} (1975) 3060;
O.~Gunnarsson, O.~Jepsen and O.~K.~Andersen:
{\it ibid.}~{\bf 27} (1983) 7144;
O.~K.~Andersen and O.~Jepsen,
Phys.~Rev.~Lett.~{\bf 53} (1984) 2571. 
%
\bibitem{sing91}
\label{singh}
D.J. Singh, D.A. Papaconstantopoulos, H. Krakauer,
B.M. Klein and W.E. Pickett:
Physica C {\bf 175} (1991) 329.
%
\bibitem{MarzariVanderbilt}
N.~Marzari and D.~Vanderbilt:
Phys.~Rev.~B~{\bf 56} (1997) 12847.
%
\bibitem{Gunnarsson}
O.~Gunnarsson, O.~K.~Andersen, O.~Jepsen and J.~Zaanen: Phys.~Rev.~B~{\bf 39} (1989) 1708;
O.~Gunnarsson, A.~V.~Postnikov and O.~K.~Andersen: \textit{ibid.}~{\bf 40} (1989) 10407;
V.~I.~Anisimov and O.~Gunnarsson:
\textit{ibid.}~{\bf 43} (1991) 7570.
%
\bibitem{anis91}
V.I. Anisimov, J. Zaanen and O.K. Andersen:
Phys. Rev. B {\bf 44} (1991) 943.
%
\bibitem{solo94}
I.V. Solovyev, P.H. Dederichs and V.I. Anisimov:
Phys. Rev. B {\bf 50} (1994) 16861.
%
\bibitem{FerdiGunnarsson}
F.~Aryasetiawan and O.~Gunnarsson:
Rep.~Prog.~Phys.~{\bf 61} (1998) 237.
%
%
\bibitem{Hedin}
L. Hedin and S. Lundqvist:
Solid State Phys.~{\bf 23} (1969) 1.
%
%
\bibitem{Hanke}
W. Hanke and L.J. Sham:
Phys.~Rev.~Lett. {\bf 33} (1974) 582.
%
\bibitem{anis97a}
V.I. Anisimov, F. Aryasetiawan and A.I. Lichtenstein:
J. Phys. Cond. Matter {\bf 9} (1997) 767.
%
\bibitem{anis97b}
V.I. Anisimov, A.I. Poteryaev, M.A. Korotin, A.O. Anokhin and G. Kotliar:
J. Phys. Cond. Matter {\bf 9} (1997) 7359.
%
\bibitem{lich98}
A.I. Lichtenstein and M.I. Katsnelson:
Phys. Rev. B {\bf 57} (1998) 6884.
%
\bibitem{held01}
K. Held, I.A. Nekrasov, N. Blumer, V.I. Anisimov and D. Vollhardt:
Int. J. Mod. Phys. B {\bf 15} (2001) 2611.
%
\bibitem{held02}
K. Held, I.A. Nekrasov, G. Keller, V. Eyert, N. Blumer, A.K. McMahan, R.T. Scalettar, T.Pruschke, V.I. Anisimov and D. Vollhardt:
{\it The LDA+DMFT Approach to Materials with Strong Electronic Correlations
Quantum Simulations of Complex Many-Body Systems: From Theory to Algorithms},
ed. J. Grotendorst, D. Marx and A. Muramatsu,
NIC Series {\bf 10} (Forschungszentrum Julich, 2002) p.175. 
%
\bibitem{kura85}
Y. Kuramoto: {\it Theory of Heavy Fermions and Valence Fluctuations},
eds. T. Kasuya and T. Saso (Springer Verlag, 1985) p.152. 
%
\bibitem{prus95}
T. Pruschke, M. Jarrell and J. K. Freericks,
Adv. Phys. {\bf 42} (1995) 187.
%
\bibitem{geor96}
A. Georges, G. Kotliar, W. Krauth and M.J. Rosenberg:
Rev. Mod. Phys. {\bf 68} (1996) 13.
%
\bibitem{Kawakami}
A. Koga, K. Inaba and N. Kawakami:
Prog. Theor. Phys. Suppl. {\bf 160} (2005) 253 and references therein.
%
\bibitem{Rice}
V.I. Anisimov, I.A. Nekrasov, D.E.Kondakov, T.M. Rice and M. Sigrist:
Eur Phys. J. B {\bf 25} (2002) 191.
%
\bibitem{Maeno}
S. Nakatsuji and Y. Maeno:
Phys. Rev. Lett. {\bf 84} (2000) 2666.
%
\bibitem{Imada2005}
M. Imada:
Phys. Rev. B {\bf 72} (2005) 075113; J. Phys. Soc. Jpn. {\bf 74} (2005) 859. 
\end{thebibliography}
\end{document}